\documentclass[final,onefignum,onetabnum]{siamart220329}

\usepackage{amsfonts,mathrsfs}
\allowdisplaybreaks
\usepackage[noend]{algpseudocode}
\usepackage{url}
\usepackage{enumitem}
\usepackage{multirow}
\usepackage{booktabs}

\algnewcommand{\LineComment}[1]{\vspace{.0625in}\State \textcolor{gray}{\texttt{\# #1}}}
\algrenewcomment[1]{\hfill\textcolor{gray}{\texttt{\# #1}}}
\newsiamthm{example}{Example}
\newsiamthm{remark}{Remark}

\renewcommand{\arraystretch}{1.2}

\begin{document}

\ifpdf
\hypersetup{
    pdftitle={Non-intrusive reduced-order models for parametric partial differential equations via data-driven operator inference},
    pdfauthor={Shane A. McQuarrie, Parisa Khodabakhshi, Karen E. Willcox},
    pdfkeywords={Parametric model reduction, operator inference, scientific machine learning, data-driven reduced model}
}
\fi

\title{Non-intrusive reduced-order models for parametric partial differential equations via data-driven operator inference}

\author{
Shane~A.~McQuarrie\thanks{Oden Institute for Computational Engineering and Sciences, University of Texas at Austin, Austin, TX, USA (\email{shanemcq@utexas.edu}, \email{kwillcox@oden.utexas.edu}).} \and Parisa~Khodabakhshi\thanks{Department of Mechanical Engineering and Mechanics, Lehigh University, Bethlehem, PA, USA (\email{pak322@lehigh.edu}).} \and Karen~E.~Willcox\footnotemark[1]
}

\headers{Affine-parametric Operator Inference}{S.~A.~McQuarrie, P.~Khodabakhshi, K.~E.~Willcox}

\maketitle

\begin{abstract}
This work formulates a new approach to reduced modeling of parameterized, time-dependent partial differential equations (PDEs). The method employs Operator Inference, a scientific machine learning framework combining data-driven learning and physics-based modeling. The parametric structure of the governing equations is embedded directly into the reduced-order model, and parameterized reduced-order operators are learned via a data-driven linear regression problem. The result is a reduced-order model that can be solved rapidly to map parameter values to approximate PDE solutions. Such parameterized reduced-order models may be used as physics-based surrogates for uncertainty quantification and inverse problems that require many forward solves of parametric PDEs. Numerical issues such as well-posedness and the need for appropriate regularization in the learning problem are considered, and an algorithm for hyperparameter selection is presented. The method is illustrated for a parametric heat equation and demonstrated for the FitzHugh--Nagumo neuron model.
\end{abstract}



\section{Introduction}
\label{sec:introduction}

Model reduction seeks to alleviate the computational burden of large-scale numerical simulations of dynamical systems by constructing reduced-order models (ROMs) that accurately capture the system dynamics, but which are much less expensive to solve than the high-fidelity models inherent in applications. The challenge is to generate ROMs from limited training data that respond well to changes in the scenario parameters that define the governing dynamics \cite{BGW2015pmorSurvery}. Such parametric ROMs are critical for enabling outer-loop applications such as design, inverse problems, optimization, and uncertainty quantification. Furthermore, as high-fidelity simulations become increasingly sophisticated and simulation data becomes more available, there is a growing need for non-intrusive model reduction methods, which aim to learn ROMs primarily from simulation data and/or outputs, as opposed to making a direct reduction of the underlying high-fidelity code that produced them \cite{GW2021learning}. Non-intrusive approaches combine data-driven learning with physics-based modeling in a way that enables both flexibility and robustness. This paper presents a framework for learning parametric ROMs in a non-intrusive fashion.

Adapting non-intrusive model reduction strategies to the parametric setting is an active area of research. One major model reduction strategy, Dynamic mode decomposition (DMD) \cite{brunton2015compressed,schmid2010dynamic}, learns a low-dimensional linear mapping based on state space data, approximating the eigenstates of the infinite-dimensional Koopman operator. The work in \cite{tezzele2020enhancing} targets parametric problems by incorporating DMD with an active subspace strategy to reduce the dimensionality of the parameter space. Methods based on the Loewner framework \cite{antoulas1986rationalInterp,antoulas2016loewnerbilinear}, another common non-intrusive approach to model reduction, build a ROM based on input-output measurements and transfer functions. Loewner methods have been generalized to parameterized linear systems by introducing additional degrees of freedom in the construction of the reduced-order transfer function to account for parametric dependencies \cite{Ionita_Antoulas_2014}. Recent work in \cite{Stuart2020neuralPDE} blends ideas from classical model reduction with deep learning to construct a mapping from parametric inputs to state outputs. Deep learning approaches to model reduction aim to benefit from the flexibility of representing the state on a low-dimensional but nonlinear manifold \cite{lee2020model}.

Equation discovery methods, in which the governing equations of a dynamical system are learned from data, share some characteristics with non-intrusive model reduction methods. One class of equation discovery approaches uses sparse regression to identify the underlying partial differential equations (PDEs) from a set of data \cite{schaeffer2017learning} or the key terms in a dynamical system within a library of potential nonlinear terms \cite{BPK2016sindy,MMKB2018sindyabrupt,rudy2019data}. In a similar vein, the work in \cite{arbabi2020linking,lee2020coarse} combines machine learning techniques with manifold learning algorithms to learn a macroscopic model for long-wavelength behavior corresponding to fine-scale measurement data. Each of these methods relies on an appropriate candidate library for the terms of the unknown equations and selects the best combination of terms based on data. In some cases where the original data is of high dimension, a dimensionality reduction technique is first applied so that the model selection occurs in a low-dimensional setting (see, e.g., \cite{BPK2016sindy,fasel2022ensemble}). Model reduction approaches also utilize the paradigm of first seeking a low-dimensional representation for the system state which, for data-driven methods, leads to a learning problem posed in a reduced space.

Operator Inference (OpInf), introduced in \cite{PW2016operatorInference}, is a non-intrusive framework for model reduction of systems with polynomial nonlinearities. As with other non-intrusive approaches, the method does not require intrusive access to source code, instead inferring the ROM solely from initial conditions, simulation snapshots, and corresponding inputs. Known governing equations motivate the form of the ROM, and the operators defining the ROM are chosen by minimizing a data-driven residual in a reduced state space. The associated learning problem is linear, dense, relatively small, and has a closed-form solution. Since its introduction, the OpInf framework has been expanded in several ways: transforming variables (lifting) to induce the requisite polynomial structure \cite{QKPW2020liftAndLearn,SKHW2020romCombustion}; approximating nonpolynomial nonlinearities via the discrete empirical interpolation method (DEIM) \cite{BGKPW2020opInfDeim}; regularizing the learning problem to enable performance on large-scale systems \cite{jain2021performance,MHW2021regOpInfCombustion}; re-projecting trajectories to exactly recover intrusive ROMs \cite{Peherstorfer2020reprojection}; and accounting for algebraic equations arising from lifting transformations \cite{KW2021_DAE}, to name a few. In these OpInf-based methods, parametric dependencies are addressed by learning separate ROMs for individual parameter samples, then interpolating either their reduced operators \cite{Peherstorfer2020reprojection,PW2016operatorInference} or their outputs \cite{KW2021_DAE}. However, interpolation in more than one or two parameter dimensions quickly becomes challenging due to the curse of dimensionality and Runge's phenomenon.

In this paper, we show that the parametric structure of the governing equations of interest can be built directly into the OpInf regression problem---circumventing the need for interpolation---if the parametric dependencies have an affine form. Affine-parametric problems have been studied frequently in the context of reduced basis methods \cite{Grepl05,RozHP08,Veroy05,Veroy03}, in which the preservation of the parametric structure by projection plays a key role \cite{BGW2015pmorSurvery}. The recent work \cite{YGBK2021} developed an OpInf framework for linear-quadratic affine-parametric problems in a semi-discrete setting, then applied the methodology successfully to the shallow water equations in non-traditional form with a single parameter. In this work, we pose the affine-parametric OpInf problem in the fully continuous setting and establish rigorous \emph{a priori} conditions for determining the well-posedness of the inference problem. These conditions can be used to inform the choice of parameter samples for training data. We extend the framework to systems of PDEs such that, different from \cite{YGBK2021}, the system-level parametric structure is exactly preserved. Leveraging this structure preservation, we propose a robust regularization strategy which can be tailored to specific PDE operators. This step is key for obtaining stable and accurate ROMs. Our approach is applied to problems with two- and four-dimensional parameter spaces, demonstrating that our framework enables ROMs for parametric PDE systems with multiple parameters. In summary, the approach presented here includes the following key contributions: 1) formulating a general OpInf framework for affine-parametric systems of PDEs in a time-continuous setting, 2) analyzing the associated well-posedness conditions, and 3) implementing a flexible regularization strategy in the numerical solution of the inference problem.

The remainder of the paper is organized as follows. \Cref{sec:methodology} establishes the general methodology; \Cref{sec:pde_systems} extends the framework to systems of PDEs; \Cref{sec:computation} details the computational aspects of solving the parametric OpInf problem; \Cref{sec:numerics} presents two numerical examples; and \Cref{sec:conclusions} concludes the paper.

\section{Non-intrusive Parametric Model Reduction}
\label{sec:methodology}
In \cref{subsec:pde_to_ode}, we show how the form of an appropriate ROM can be determined directly from the form of certain PDEs; \cref{subsec:operator_inference} presents the OpInf approach for learning such ROMs from data and system structure. We also introduce a heat equation example for which numerical results are reported in \cref{subsec:numerical_heat}.

\subsection{Projection-based Reduced-order Models of Parametric PDEs}
\label{subsec:pde_to_ode}
We target systems governed by parametric PDEs that are polynomial in state, which includes linear PDEs as well as a large class of nonlinear PDEs. Let $\Omega \subset \mathbb{R}^{d_x}$ be an open, bounded set with Lipschitz continuous boundary $\partial\Omega = \Gamma \cup (\partial\Omega \setminus \Gamma)$ and outward-pointing normal $\boldsymbol{\eta}\in\mathbb{R}^{d_{x}}$. For the time domain $[t_0,t_f]\subset\mathbb{R}$ and the parameter domain $\mathcal{P}\subset\mathbb{R}^{d_\mu}$, we consider the initial/boundary-value problem
\begin{subequations}
\begin{align}
    \label{eq:pde-single}
    \frac{\partial u}{\partial t}
    &= \mathcal{F}(u;\mu),
    &
    x &\in \Omega,\
    t \in (t_0,t_f],\
    \mu \in \mathcal{P},
    \\
    \label{eq:pde-initial}
    u(x, t_0; \mu) &= u_0(x;\mu),
    &
    x &\in \Omega,\
    \mu \in \mathcal{P},
    \\
    \label{eq:pde-bcs}
    u(x, t; \mu) &= 0,
    &
    x &\in \Gamma,\
    t\in[t_0,t_f],\
    \mu \in \mathcal{P},
    \\
    \label{eq:pde-bcs-neumann}
    \boldsymbol{\eta}\cdot\nabla_x u(x, t; \mu) &= 0,
    &
    x &\in \partial\Omega \setminus \Gamma,\
    t\in[t_0,t_f],\
    \mu \in \mathcal{P},
\end{align}
\end{subequations}
where the unknown state variable $u(\cdot,t;\mu)$ is contained in a separable Hilbert space $\mathcal{V}$ of real-valued functions satisfying the boundary conditions \cref{eq:pde-bcs}--\cref{eq:pde-bcs-neumann} with dual space $\mathcal{V}^{*}$, and $\mathcal{F}:\mathcal{V}\times\mathcal{P}\to\mathcal{V}^{*}$ is a spatial differential operator that depends on the free parameter $\mu\in\mathcal{P}$. We assume \cref{eq:pde-single}--\cref{eq:pde-bcs-neumann} has a unique solution $u$ in at least the weak sense, meaning
\begin{align}
    \label{eq:weak-solution}
    \left\langle v, \frac{\partial u}{\partial t} \right\rangle
    = \left\langle v, \mathcal{F}(u;\mu) \right\rangle
    \qquad
    \textrm{for all}
    \qquad
    v\in\mathcal{V},\
    t\in[t_0,t_f],\
    \mu\in\mathcal{P},
\end{align}
where $\langle\cdot,\cdot\rangle$ is the duality pairing of $\mathcal{V}$ with $\mathcal{V}^{*}$.

We consider the setting in which $\mathcal{F}$ has a polynomial structure with respect to the state $u$ and its spatial derivatives. Many PDEs enjoy this structure or can be written in this form through a change of variables \cite{QKPW2020liftAndLearn}. For brevity we consider a quadratic form, but higher-order (e.g., cubic) terms may also be included, as we will see later in \Cref{example:fh-n}. Specifically, suppose
\begin{subequations}
\begin{align}
    \label{eq:pde-polynomial}
    \mathcal{F}(u;\mu)
    = \mathcal{C}(\mu)
    + \mathcal{A}(u;\mu)
    + \mathcal{H}(u,u;\mu),
\end{align}
where $\mathcal{C}:\mathcal{P}\to\mathcal{V}^{*}$, and where $\mathcal{A}:\mathcal{V}\times\mathcal{P}\to\mathcal{V}^{*}$ and $\mathcal{H}:\mathcal{V}\times\mathcal{V}\times\mathcal{P}\to\mathcal{V}^{*}$ are linear in each of their state arguments. Furthermore, assume that the operators $\mathcal{C}$, $\mathcal{A}$, and $\mathcal{H}$ exhibit the following affine decompositions with respect to the parameter $\mu$:
\begin{gather}
    \label{eq:affine-op}
    \begin{gathered}
    \mathcal{C}(\mu) = \sum_{p=1}^{q_c}\theta_{c}^{(p)}(\mu)\mathcal{C}^{(p)},
    \qquad
    \mathcal{A}(u;\mu) = \sum_{p=1}^{q_A}\theta_{A}^{(p)}(\mu)\mathcal{A}^{(p)}(u),
    \\
    \mathcal{H}(u,v;\mu)
    = \sum_{p=1}^{q_H}\theta_{H}^{(p)}(\mu)\mathcal{H}^{(p)}(u,v),
    \end{gathered}
\end{gather}
\end{subequations}
where $\mathcal{C}^{(p)}\in\mathcal{V}^{*}$, $\mathcal{A}^{(p)}:\mathcal{V}\to\mathcal{V}^{*}$, $\mathcal{H}^{(p)}:\mathcal{V}\times\mathcal{V}\to\mathcal{V}^{*}$, and the scalar-valued functions $\theta_{c}^{(p)},\theta_{A}^{(p)},\theta_{H}^{(p)}:\mathcal{P}\to\mathbb{R}$ are such that the sets $\{\theta_{c}^{(p)}\}_{p=1}^{q_c}$, $\{\theta_{A}^{(p)}\}_{p=1}^{q_A}$, and $\{\theta_{H}^{(p)}\}_{p=1}^{q_{H}}$ are each linearly independent. Note that the operators $\mathcal{C}^{(p)}$, $\mathcal{A}^{(p)}$, and $\mathcal{H}^{(p)}$ are independent of the parameter $\mu$. This work considers the class of problems where the affine-parametric coefficient functions ($\theta_{c}^{(p)},\theta_{A}^{(p)},\theta_{H}^{(p)}$) are known. Such a structure may occur naturally in the known governing equations, be induced via approximation with the empirical interpolation method \cite{BMNP2004eim}, or in some cases discovered from data via sparse regression \cite{rudy2019data,schaeffer2017learning}. We leave for future work the possibility of simultaneously inferring the structure \cref{eq:affine-op} and constructing a ROM from data; see \cite{lieberman2010parameter} for an example of simultaneous reduction in both the parameter and state space.

A projection-based ROM of \cref{eq:pde-single}--\cref{eq:pde-bcs-neumann} with $\mathcal{F}$ as in \cref{eq:pde-polynomial}--\cref{eq:affine-op} retains the affine-parametric polynomial structure of the system \cite{BGW2015pmorSurvery,GPT1999vortexsheddingPOD}. Let $\{v_{j}\}_{j=1}^{\infty}\subset\mathcal{V}$ be an orthonormal set such that the solution $u$ may be expressed with the expansion
\begin{align}
    \label{eq:basis-expansion}
    u(x,t;\mu)
    = \sum_{j=1}^{\infty} \hat{u}_{j}(t;\mu)v_{j}(x).
\end{align}
Since $\langle v_i, v_j \rangle = \delta_{ij}$, the coefficients satisfy $\hat{u}_{j}(t;\mu) = \left\langle v_j, u(\cdot,t;\mu) \right\rangle$. A reduced model with $r \in \mathbb{N}$ degrees of freedom consists of time evolution equations for the coefficients $\hat{u}_{1}(t;\mu),\ldots,\hat{u}_{r}(t;\mu)$; the approximate ROM solution $\breve{u}$ of \cref{eq:pde-single}--\cref{eq:pde-bcs-neumann} is then given by the sum \cref{eq:basis-expansion}, truncated to $r$ terms:
\begin{align}
    \label{eq:truncation-projection}
    \breve{u}(x,t;\mu)
    = \sum_{j=1}^{r} \hat{u}_{j}(t;\mu)v_{j}(x).
\end{align}
Note that $\breve{u}$ is confined to the finite-dimensional subspace $\textrm{span}(\{v_1,\ldots,v_r\})\subset\mathcal{V}$. By substituting $\breve{u}$ for $u$ in \cref{eq:weak-solution} with test function $v=v_{i}$, and using the form of $\mathcal{F}$ from \cref{eq:pde-polynomial}--\cref{eq:affine-op}, we obtain
\begin{align}
    \label{eq:inner_product}
    \begin{aligned}
    \frac{\textup{d} \hat{u}_{i}}{\textup{d} t}
    = \sum_{p=1}^{q_{c}}\theta_{c}^{(p)}(\mu)
        \left\langle v_i, \mathcal{C}^{(p)} \right\rangle
    &+ \sum_{p=1}^{q_{A}}\theta_{A}^{(p)}(\mu)\sum_{j=1}^{r}
        \left\langle v_i, \mathcal{A}^{(p)}\left(v_{j}\right)\right\rangle
        \hat{u}_{j}
    \\
    &+ \sum_{p=1}^{q_{H}}\theta_{H}^{(p)}(\mu)\sum_{j=1}^{r}\sum_{k=1}^{r}
        \left\langle v_i,\mathcal{H}^{(p)}\left(v_{j},v_{k}\right)\right\rangle
        \hat{u}_{j} \hat{u}_{k}.
    \end{aligned}
\end{align}
Collecting \cref{eq:inner_product} for $i=1,\ldots,r$ yields a uniquely-defined system of ordinary differential equations (ODEs) with state vector $
\widehat{\mathbf{u}}(t;\mu)
    = [
    \hat{u}_{1}(t;\mu)~\cdots~\hat{u}_{r}(t;\mu)
    ]^\top
\in \mathbb{R}^{r},
$
\begin{subequations}
\begin{align}
    \label{eq:ode-affine}
    \begin{aligned}
    \frac{\textup{d}}{\textup{d}t}\widehat{\mathbf{u}}(t;\mu)
    = \left(\sum_{p=1}^{q_c}\theta_{c}^{(p)}(\mu)\widehat{\mathbf{c}}^{(p)}\right)
    &+ \left(\sum_{p=1}^{q_A}\theta_{A}^{(p)}(\mu)\widehat{\mathbf{A}}^{(p)}\right)\widehat{\mathbf{u}}(t;\mu)
    \\
    &+ \left(\sum_{p=1}^{q_H}\theta_{H}^{(p)}(\mu)\widehat{\mathbf{H}}^{(p)}\right)\left(\widehat{\mathbf{u}}(t;\mu)\, \widehat{\odot}\, \widehat{\mathbf{u}}(t;\mu)\right),
    \end{aligned}
    \\
    \label{eq:ode-initial-condition}
    \widehat{\mathbf{u}}(t_0;\mu)
    = \left[\begin{array}{ccc}
    \left\langle v_{1}, u_0(\mu) \right\rangle & \cdots & \left\langle v_{r}, u_0(\mu) \right\rangle
    \end{array}\right]^\top \in \mathbb{R}^{r},
\end{align}
\end{subequations}
where $\widehat{\mathbf{c}}^{(p)}\in\mathbb{R}^{r}$, $\widehat{\mathbf{A}}^{(p)}\in\mathbb{R}^{r \times r}$, $\widehat{\mathbf{H}}^{(p)}\in\mathbb{R}^{r \times \binom{r+1}{2}}$, and $\widehat{\odot}$ denotes a compact Khatri-Rao product, i.e., $\mathbf{u}\,\widehat{\odot}\,\mathbf{u}$ extracts the unique components of the Khatri-Rao product of $\mathbf{u}$ with itself (see \cref{appendix:kronecker}). Note that \cref{eq:ode-affine} and \cref{eq:pde-polynomial}--\cref{eq:affine-op} are both quadratic in their respective state and affine-parametric with respect to $\mu$, with the same functions $\theta_{c}^{(p)}$, $\theta_{A}^{(p)}$, and $\theta_{H}^{(p)}$ in the affine expansions.

The system \cref{eq:ode-affine}--\cref{eq:ode-initial-condition} is a ROM for \cref{eq:pde-single}--\cref{eq:pde-bcs-neumann} in which the boundary conditions are embedded through the basis functions. The quality of the ROM as a surrogate for \cref{eq:pde-single}--\cref{eq:pde-bcs-neumann} depends heavily on the basis $\{v_{j}\}_{j=1}^{r}$, but the \emph{form} of the equations is the same for any choice of basis. In \cref{subsec:operator_inference}, we leverage this property to develop a procedure for learning such a model without explicitly evaluating the terms in \cref{eq:inner_product}.

\begin{remark}
As written above, \cref{eq:ode-affine} is the most general form of a quadratic ODE with affine-parametric structure as in \cref{eq:affine-op}. However, as we will see in examples, the constant, linear, and/or quadratic terms are not always present in practice, and when they are, the number of terms in the affine expansions (i.e., $q_c$, $q_A$, and $q_H$) tends to be small. In other words, projection-based ROMs of the form \cref{eq:ode-affine} have the same number of nonzero terms as the original PDE \cref{eq:pde-single}.
\end{remark}

\begin{example}[Heat Equation]
\label{example:heat-1}
Consider the one-dimensional spatial domain $\Omega = (0,1) \subset \mathbb{R}$ and the parameter domain $\mathcal{P} = [.01,2.5]\times[0.1,2.5]\subset\mathbb{R}^{2}$. For a fixed $\bar{x}\in\Omega$, let $\chi_{[0,\bar{x})}$ and $\chi_{[\bar{x},1]}$ be indicator functions over $\Omega$,
\begin{align*}
    \chi_{[0,\bar{x})}(x)
    &= \begin{cases}
        1, & 0 < x < \bar{x}, \\
        0, & \bar{x} \le x < 1,
    \end{cases}
    &
    \chi_{[\bar{x},1]}(x)
    &= \begin{cases}
        0, & 0 < x < \bar{x}, \\
        1, & \bar{x} \le x < 1.
    \end{cases}
\end{align*}
The following equation models the diffusion of heat through a one-dimensional rod composed of two materials with independent thermal diffusivities $\mu = (\alpha,\beta)\in\mathcal{P}$, with the temperature prescribed at each end of the rod:
\begin{subequations}
\begin{align}
    \label{eq:heat-pde}
    \frac{\partial u}{\partial t}
    &= \left(\alpha \chi_{[0,\bar{x})} + \beta \chi_{[\bar{x},1]}\right) \frac{\partial^2 u}{\partial x^2}
    &
    x &\in \Omega,\
    t \in (t_0,t_f],\
    \mu\in\mathcal{P},
    \\
    \label{eq:heat-init}
    u(x, t_0; \mu) &= u_0(x;\mu),
    &
    x &\in \Omega,\
    \mu\in\mathcal{P},
    \\
    \label{eq:heat-bcs}
    u(0, t; \mu) &= u(1, t; \mu) = 0,
    &
    t &\in [t_0,t_f],\ \mu\in\mathcal{P}.
\end{align}
\end{subequations}
The underlying Hilbert space containing the state $u$ is $\mathcal{V} = H^2(\Omega)\cap H^{1}_{0}(\Omega)$, the set of twice weakly differentiable functions satisfying the homogeneous Dirichlet boundary conditions \cref{eq:heat-bcs}. In the language of \cref{eq:pde-polynomial}--\cref{eq:affine-op}, $\mathcal{C}\equiv\mathcal{H}\equiv 0$ and
\begin{align*}
    \mathcal{A}(u;\mu)
    &= \theta_{A}^{(1)}(\mu)\mathcal{A}^{(1)}(u) + \theta_{A}^{(2)}(\mu)\mathcal{A}^{(2)}(u)
\end{align*}
with $\theta_{A}^{(1)}(\mu)=\alpha$, $\theta_{A}^{(2)}(\mu)=\beta$, $\mathcal{A}^{(1)}(u) = \chi_{[0,\bar{x})}\frac{\partial^2 u}{\partial x^2}$, and $\mathcal{A}^{(2)}(u) = \chi_{[\bar{x},1]}\frac{\partial^2 u}{\partial x^2}$. A projection-based ROM for this problem as in \cref{eq:ode-affine} has the form
\begin{align}
    \label{eq:heat-ode}
    \frac{\mathrm{d}}{\mathrm{d}t}\widehat{\mathbf{u}}(t;\mu)
    &= \left(
        \alpha\widehat{\mathbf{A}}^{(1)} + \beta\widehat{\mathbf{A}}^{(2)}
    \right)\widehat{\mathbf{u}}(t;\mu),
\end{align}
with the initial condition as in \cref{eq:ode-initial-condition}.
\end{example}

\subsection{Affine Operator Inference for PDEs}
\label{subsec:operator_inference}
Constructing $\widehat{\mathbf{c}}^{(p)}$, $\widehat{\mathbf{A}}^{(p)}$, and $\widehat{\mathbf{H}}^{(p)}$ in the ODE system \cref{eq:ode-affine} by evaluating the terms of \cref{eq:inner_product} is an inherently intrusive process, requiring explicit access to the differential operators $\mathcal{C}^{(p)}$, $\mathcal{A}^{(p)}$, and $\mathcal{H}^{(p)}$. In practice, this is not always favorable. For example, when the training data comes from commercial software, access to the details of the discretization might be limited. In addition, variable transformations used to induce the polynomial structure \cref{eq:pde-polynomial} make these differential operators computationally inaccessible \cite{KW2021_DAE,MHW2021regOpInfCombustion,QKPW2020liftAndLearn}. In this section, we pose an affine-parametric inference procedure to \emph{learn} a ROM through a data-driven optimization problem given only samples of the solution $u$, a finite basis $\{v_{j}\}_{j=1}^{r}$ (which can be computed from the solution data, as we will discuss in \Cref{sec:computation}), and knowledge of the governing equations \cref{eq:pde-single}--\cref{eq:pde-bcs} and the parametric structure \cref{eq:pde-polynomial}--\cref{eq:affine-op}. We do not rely on evaluations of the differential operators as in \cref{eq:inner_product}, but rather on the induced ODE structure \cref{eq:ode-affine}--\cref{eq:ode-initial-condition}. When restricted to linear-quadratic PDEs of a single variable, the formulation in this section is equivalent to the approach in \cite{YGBK2021}; \Cref{sec:pde_systems} generalizes our method to systems of PDEs. Although we focus on the quadratic form motivated by \cref{eq:pde-polynomial}, the theory presented here applies to affine-parametric polynomial systems of arbitrary order.

Consider \cref{eq:ode-affine} with fixed, known integers $q_{c},q_{A},q_{H} \ge 0$ and affine coefficient functions $\theta = \{\theta_{c}^{(1)},\ldots,\theta_{H}^{(q_H)}\}$, which define the polynomial and affine-parametric structure of the system. Define
\begin{subequations}
\begin{align}
    \label{eq:operator-family}
    \begin{aligned}
    \frac{\textup{d}}{\textup{d}t}\widehat{\mathbf{u}}(t;\mu)
    = \mathbf{F}(\widehat{\mathbf{O}};\widehat{\mathbf{u}},t,\theta,\mu)
    = \left(\sum_{p=1}^{q_c}\theta_{c}^{(p)}(\mu)\widehat{\mathbf{c}}^{(p)}\right)
    + \left(\sum_{p=1}^{q_A}\theta_{A}^{(p)}(\mu)\widehat{\mathbf{A}}^{(p)}\right)\widehat{\mathbf{u}}(t;\mu)\\
    + \left(\sum_{p=1}^{q_H}\theta_{H}^{(p)}(\mu)\widehat{\mathbf{H}}^{(p)}\right)\left(\widehat{\mathbf{u}}(t;\mu)\,\widehat{\odot}\,\widehat{\mathbf{u}}(t;\mu)\right),
    \end{aligned}
\end{align}
where the as yet unknown \emph{operator matrix} $\widehat{\mathbf{O}}$ is the concatenation
\begin{align}
    \label{eq:operator-matrix}
    \widehat{\mathbf{O}} = \left[\begin{array}{c|c|c}
        \widehat{\mathbf{c}}^{(1)}\ \cdots\ \widehat{\mathbf{c}}^{(q_c)} &
        \widehat{\mathbf{A}}^{(1)}\ \cdots\ \widehat{\mathbf{A}}^{(q_A)} &
        \widehat{\mathbf{H}}^{(1)}\ \cdots\ \widehat{\mathbf{H}}^{(q_H)}
    \end{array}\right]
    \in\mathbb{R}^{r\times q(r)},
\end{align}
\end{subequations}
with column dimension $q(r) = q_c + q_A r + q_H \binom{r+1}{2}$. Equipped with the initial condition \cref{eq:ode-initial-condition}, \cref{eq:operator-family}--\cref{eq:operator-matrix} describe the family of ROMs with the same polynomial and parametric form as \cref{eq:ode-affine}, with particular realizations determined by the operator matrix $\widehat{\mathbf{O}}$. The goal is to choose $\widehat{\mathbf{O}}$ such that the corresponding ROM accurately captures the dynamics of the governing PDE \cref{eq:pde-single}--\cref{eq:pde-bcs-neumann} for all $\mu\in\mathcal{P}$.

We learn $\widehat{\mathbf{O}}$ by solving a data-driven least-squares regression problem. Suppose we can sample the solution of \cref{eq:pde-single}--\cref{eq:pde-bcs-neumann} at $s$ parameter values $\{\mu_{i}\}_{i=1}^{s}\subset\mathcal{P}$ and $K$ times $\{t_{j}\}_{j=1}^{K}\subset[t_0,t_f]$. We define the loss function $\mathcal{L}:\mathbb{R}^{r\times q(r)}\to\mathbb{R}$ associated with these solution samples to be the sum of the residuals of \cref{eq:operator-family},
\begin{align}
    \label{eq:opinf-loss}
    \mathcal{L}(\widehat{\mathbf{O}})
    &=
    \sum_{i=1}^{s}\sum_{j=1}^{K}\left\|
        \mathbf{F}(\widehat{\mathbf{O}}; \widehat{\mathbf{u}}, t_{j}, \theta, \mu_{i})
        - \frac{\textup{d}}{\textup{d} t}\widehat{\mathbf{u}}(t;\mu_{i})\Bigr|_{t=t_{j}}
    \right\|_2^2,
\end{align}
where
$
    \widehat{\mathbf{u}}(t;\mu)
    = [
        \langle v_1, u(\cdot,t;\mu)\rangle
        ~\cdots~
        \langle v_r, u(\cdot,t;\mu)\rangle
    ]^\top \in \mathbb{R}^{r}
$
as before, and with $\mathbf{F}$ and $\widehat{\mathbf{O}}$ as in \cref{eq:operator-family}--\cref{eq:operator-matrix}. To write the minimization of \cref{eq:opinf-loss} in a standard form, define the matrices
\begin{align*}
    \widehat{\mathbf{U}}(\mu_{i}) &= \left[\begin{array}{ccc}
        \widehat{\mathbf{u}}(t_{1};\mu_i) & \cdots & \widehat{\mathbf{u}}(t_{K};\mu_i)
    \end{array}\right] \in \mathbb{R}^{r\times K},
\\
    \dot{\widehat{\mathbf{U}}}(\mu_{i}) &= \left[\begin{array}{ccc}
        \frac{\textup{d}}{\textup{d}t}\widehat{\mathbf{u}}(t;\mu_i)\Bigr|_{t=t_{1}} & \cdots & \frac{\textup{d}}{\textup{d}t}\widehat{\mathbf{u}}(t;\mu_i)\Bigr|_{t=t_{K}}
    \end{array}\right] \in \mathbb{R}^{r\times K},
\end{align*}
which group the projected solution and the associated time derivatives for each parameter sample, and define the row vectors
\begin{align*}
    \boldsymbol{\theta}_{c}(\mu_{i}) &= \left[\begin{array}{ccc}
        \theta_{c}^{(1)}(\mu_{i}) & \cdots & \theta_{c}^{(q_c)}(\mu_{i})
    \end{array}\right]\in\mathbb{R}^{1 \times q_c},
    \\
    \boldsymbol{\theta}_{A}(\mu_{i}) &= \left[\begin{array}{ccc}
        \theta_{A}^{(1)}(\mu_{i}) & \cdots & \theta_{A}^{(q_A)}(\mu_{i})
    \end{array}\right]\in\mathbb{R}^{1 \times q_A},
    \\
    \boldsymbol{\theta}_{H}(\mu_{i}) &= \left[\begin{array}{ccc}
        \theta_{H}^{(1)}(\mu_{i}) & \cdots & \theta_{H}^{(q_H)}(\mu_{i})
    \end{array}\right]\in\mathbb{R}^{1 \times q_H},
\end{align*}
which encode the affine-parametric dependencies of $\mathcal{C}$, $\mathcal{A}$, and $\mathcal{H}$, respectively. Finally, define the \emph{data matrix}
\begin{align}
    \label{eq:data_matrix}
    \begin{aligned}
    \mathbf{D}
    &= \left[\begin{array}{c|c|c}
        \makebox[1.96cm]{$\mathbf{D}_{c}$} &
        \makebox[2.70cm]{$\mathbf{D}_{A}$} &
        \makebox[4.55cm]{$\mathbf{D}_{H}$}
    \end{array}\right]
    \\
    &= \left[\begin{array}{c|c|c}
        \boldsymbol{\theta}_c(\mu_{1})\otimes\mathbf{1}_{K} &
        \boldsymbol{\theta}_A(\mu_{1})\otimes \widehat{\mathbf{U}}(\mu_{1})^\top &
        \boldsymbol{\theta}_H(\mu_{1})\otimes\left(\widehat{\mathbf{U}}(\mu_{1}) \,\widehat{\odot}\, \widehat{\mathbf{U}}(\mu_{1})\right)^\top
        \\
        \vdots & \vdots & \vdots \\
        \boldsymbol{\theta}_c(\mu_{s})\otimes\mathbf{1}_{K} &
        \boldsymbol{\theta}_A(\mu_{s})\otimes \widehat{\mathbf{U}}(\mu_{s})^\top &
        \boldsymbol{\theta}_H(\mu_{s})\otimes\left(\widehat{\mathbf{U}}(\mu_{s})\,\widehat{\odot}\,\widehat{\mathbf{U}}(\mu_{s})\right)^\top
    \end{array}\right], \end{aligned}
\end{align}
where $\mathbf{1}_{K}\in \mathbb{R}^{K}$ is a column vector of unity of length $K$. We then have
\begin{align}
    \label{eq:opinf-standard}
    \text{pOpInf:}
    \qquad
    \min_{\widehat{\mathbf{O}}}
    \mathcal{L}(\widehat{\mathbf{O}})
    =
    \min_{\widehat{\mathbf{O}}}
    \left\|
        \mathbf{D}\widehat{\mathbf{O}}^\top - \mathbf{R}^\top
    \right\|_{F}^2,
\end{align}
where $\mathbf{D} \in \mathbb{R}^{sK\times q\left(r\right)}$, $\widehat{\mathbf{O}}\in \mathbb{R}^{r\times q\left(r\right)}$, and
$
    \mathbf{R} = [
        \dot{\widehat{\mathbf{U}}}(\mu_{1})
        ~\cdots~
        \dot{\widehat{\mathbf{U}}}(\mu_{s})
    ] \in\mathbb{R}^{r \times sK}.
$
We call \cref{eq:opinf-standard} the \emph{affine-parametric Operator Inference problem} (pOpInf). Note that \cref{eq:opinf-standard} is a linear least-squares problem which decouples by the columns of $\widehat{\mathbf{O}}^\top$, meaning the dynamics for each $\hat{u}_{i}$ are learned independently \cite{PW2016operatorInference}.

If the data matrix $\mathbf{D}$ has full column rank, then \cref{eq:opinf-standard} has a unique closed-form solution \cite{bjorck1996numerical,golub2013matrix,trefethen1997numerical}. However, $\mathbf{D}$ is susceptible to rank deficiencies due to its Kronecker block structure. \Cref{thm:rank-deficiencies} establishes necessary conditions for the well-posedness of the pOpInf problem \cref{eq:opinf-standard} and sufficient conditions in the monomial case (e.g., $\mathcal{C} = \mathcal{A} = 0$ but $\mathcal{H} \neq 0$ so that $\mathbf{D} = \mathbf{D}_{H}$).

\begin{theorem}
\label{thm:rank-deficiencies}
Let $\Theta_{c}\in\mathbb{R}^{s\times q_{c}}$, $\Theta_{A}\in\mathbb{R}^{s\times q_{A}}$, and  $\Theta_{H}\in\mathbb{R}^{s\times q_{H}}$ be the matrices with entries
\begin{align}
    \label{eq:Theta-matrices}
    [\Theta_{c}]_{ij}
    &=    \theta_{c}^{(j)}(\mu_{i}),
    &
    [\Theta_{A}]_{ij}
    &=    \theta_{A}^{(j)}(\mu_{i}),
    &
    [\Theta_{H}]_{ij}
    &=    \theta_{H}^{(j)}(\mu_{i}),
\end{align}
that is, the $i$th row of $\Theta_{c}$ is $\boldsymbol{\theta}_{c}(\mu_{i})$, the $i$th row of $\Theta_{A}$ is $\boldsymbol{\theta}_{A}(\mu_{i})$, and the $i$th row of $\Theta_{H}$ is $\boldsymbol{\theta}_{H}(\mu_{i})$. If any of $\Theta_c$, $\Theta_{A}$, or $\Theta_{H}$ do not have full column rank, then the data matrix $\mathbf{D} = [\mathbf{D}_{c}~\mathbf{D}_{A}~\mathbf{D}_{H}]$ defined in \cref{eq:data_matrix} is rank deficient. Furthermore, if either of the block matrices
\begin{align*}
    \widehat{\mathbf{U}}_{A}
    &= \left[\begin{array}{c}
        \widehat{\mathbf{U}}(\mu_{1})^\top
        \\ \vdots \\
        \widehat{\mathbf{U}}(\mu_{s})^\top
    \end{array}\right],
    &
    \widehat{\mathbf{U}}_{H}
    &= \left[\begin{array}{c}
        \left(\widehat{\mathbf{U}}(\mu_{1})\,\widehat{\odot}\,\widehat{\mathbf{U}}(\mu_{1})\right)^\top
        \\ \vdots \\
        \left(\widehat{\mathbf{U}}(\mu_{s})\,\widehat{\odot}\,\widehat{\mathbf{U}}(\mu_{s})\right)^\top
    \end{array}\right]
\end{align*}
do not have full column rank, then neither does $\mathbf{D}$. On the other hand, if $\Theta_{A}$ and each $\widehat{\mathbf{U}}(\mu_{i})^\top$ ($i=1,\ldots,s$) have full column rank, then so does $\mathbf{D}_{A}$; and if $\Theta_{H}$ and each $(\widehat{\mathbf{U}}(\mu_{i})\,\widehat{\odot}\,\widehat{\mathbf{U}}(\mu_{i}))^\top$ have full column rank, then so does $\mathbf{D}_{H}$.
\begin{proof}
Note that $\Theta_{c} = \mathbf{D}_{c}$, the leftmost block of $\mathbf{D}$. Therefore, if $\Theta_{c}$ does not have full column rank, then $\mathbf{D}$ has linearly dependent columns. \Cref{lemma:kronecker-structure} gives the result for the remaining cases: if $\Theta_{A}$ or $\widehat{\mathbf{U}}_{A}$ do not have full column rank, apply \cref{lemma:kronecker-structure} with $\mathbf{y}_{i} = \boldsymbol{\theta}_{A}(\mu_{i})^\top$, $\mathbf{Z}_{i} = \widehat{\mathbf{U}}(\mu_{i})^{\top}$, and $\mathbf{W} = \mathbf{D}_{A}$ to show $\mathbf{D}_{A}$ is rank deficient; a similar argument holds for $\Theta_{H}$ and $\widehat{\mathbf{U}}_{H}$, showing $\mathbf{D}_{H}$ is rank deficient. The results to the converse also hold by \cref{lemma:kronecker-structure}.
\end{proof}
\end{theorem}

Selecting appropriate parameter samples is an important issue for all parametric model reduction methods \cite{BGW2015pmorSurvery}. \Cref{thm:rank-deficiencies} provides numerically relevant guidance for parameter selection in the context of \cref{eq:opinf-standard}: for a set of parameter values $\{\mu_{i}\}_{i=1}^{s}$, the (small) matrices $\Theta_{c}$, $\Theta_{A}$, and $\Theta_{H}$ can be explicitly formed and checked for rank deficiencies without any information about the solution trajectories. If any of the matrices are unsatisfactorily conditioned, different parameter values must be chosen at which to sample the solution. This is especially relevant for settings in which only a few solution trajectories can be afforded, i.e., $s$ is relatively small. Note that if $s < \max\{q_{c},q_{A},q_{H}\}$, then the problem is guaranteed to be ill-posed. \Cref{sec:computation} further addresses numerical issues that may arise from other sources of poor conditioning.

\begin{example}[Heat Equation]
\label{example:heat-2}
Recall the problem introduced in \cref{example:heat-1}. Given parameter samples $\{\mu_{i} = (\alpha_{i},\beta_{i})\}_{i=1}^{s}$, we learn a ROM of the form \cref{eq:heat-ode} by solving the pOpInf problem \cref{eq:opinf-standard} with operator and data matrices
\begin{align*}
    \widehat{\mathbf{O}} &= \left[\begin{array}{cc}
        \widehat{\mathbf{A}}^{(1)} & \widehat{\mathbf{A}}^{(2)}
    \end{array}\right] \in \mathbb{R}^{r \times 2r},
    &
    \mathbf{D} &= \left[\begin{array}{cc}
        \alpha_{1} \widehat{\mathbf{U}}(\mu_1)^\top & \beta_{1} \widehat{\mathbf{U}}(\mu_1)^\top \\
        \vdots & \vdots \\
        \alpha_{s} \widehat{\mathbf{U}}(\mu_s)^\top & \beta_{s} \widehat{\mathbf{U}}(\mu_s)^\top \\
    \end{array}\right] \in \mathbb{R}^{sK \times 2r}.
\end{align*}
Per \cref{thm:rank-deficiencies}, the rank of the data matrix $\mathbf{D}$ depends on each $\widehat{\mathbf{U}}(\mu_{i})$ and the matrix
\begin{align*}
    \Theta_{A}
= \left[\begin{array}{cc}
        \theta_{A}^{(1)}(\mu_1) & \theta_{A}^{(2)}(\mu_1) \\
        \vdots & \vdots \\
        \theta_{A}^{(1)}(\mu_s) & \theta_{A}^{(2)}(\mu_s)
    \end{array}\right]
    = \left[\begin{array}{cc}
        \alpha_{1} & \beta_{1} \\
        \vdots & \vdots \\
        \alpha_{s} & \beta_{s}
    \end{array}\right]\in\mathbb{R}^{s\times 2}.
\end{align*}
In particular, if $\Theta_{A}$ does not have full column rank, then the data matrix $\mathbf{D}$ will be rank deficient. Our goal, then, is to choose $s \ge 2$ parameter samples such that $\Theta_{A}$ has a small condition number.
\end{example}

Reduced-order models learned through pOpInf and those constructed via intrusive projection, i.e., by explicitly evaluating \cref{eq:inner_product}, are related in the following sense.

\begin{theorem}
\label{thm:existence-uniqueness}
Let $u$ be the unique solution of \cref{eq:pde-single}--\cref{eq:pde-bcs-neumann}, where the differential operator $\mathcal{F}$ has the quadratic, affine-parametric form described in \cref{eq:pde-polynomial}--\cref{eq:affine-op}. Suppose there exists an orthonormal set $\{v_{j}\}_{j=1}^{r}\subset\mathcal{V}$, and let $\{\mu_{i}\}_{i=1}^{s}\subset\mathcal{P}$ be a finite set of parameter samples. Define the loss function
\begin{align}
    \label{eq:loss-timecontinuous}
    \mathscr{L}(\widehat{\mathbf{O}}) =
    \sum_{i=1}^{s}
     \int_{t_0}^{t_f}
      \left\|
       \mathbf{F}\left(
         \widehat{\mathbf{O}};\widehat{\mathbf{u}},t,\theta,\mu_{i}
        \right)
       - \frac{\textup{d}}{\textup{d}t}\widehat{\mathbf{u}}(t;\mu_{i})
      \right\|_{2}^{2}
     \:dt,
\end{align}
where
$
    \widehat{\mathbf{u}}(t;\mu)
    = [
        \langle v_1, u(\cdot,t;\mu)\rangle
        ~\cdots~
        \langle v_r, u(\cdot,t;\mu)\rangle
    ]^\top \in \mathbb{R}^{r},
$
and with $\mathbf{F}$ and $\widehat{\mathbf{O}}$ given by \cref{eq:operator-family}--\cref{eq:operator-matrix}. Consider the following conditions.
\begin{enumerate}
\item There exist functions $\hat{u}_{j}:[t_0,t_f]\times\mathcal{P}\to\mathbb{R}$, $j=1,\ldots,r$, such that the finite sum representation \cref{eq:truncation-projection} is exact for all $x\in\Omega$, $t\in[t_0,t_f]$, and $\mu \in \{\mu_{i}\}_{i=1}^{s}$, i.e., there is no truncation error at the sample parameter values.

\item For $i=1,\ldots,s$, there exist times $\{\tau_{i,j}\}_{j=1}^{K}\subset [t_0,t_f]$, $K = 1 + r + \binom{r+1}{2}$, such that the matrix
\begin{align*}
    \widetilde{\mathbf{D}}(\mu_i) = \left[\begin{array}{ccc}
        \mathbf{1}_{K} &
        \widetilde{\mathbf{U}}(\mu_{i})^\top &
        \left(\widetilde{\mathbf{U}}(\mu_{i})\,\widehat{\odot}\,\widetilde{\mathbf{U}}(\mu_{i})\right)^\top
    \end{array}\right] \in \mathbb{R}^{K \times K}
\end{align*}
is invertible, where
$
\widetilde{\mathbf{U}}(\mu_{i})
    = [
        \widehat{\mathbf{u}}(\tau_{i,1};\mu_i)~\cdots~\widehat{\mathbf{u}}(\tau_{i,K};\mu_i)
    ]
    \in\mathbb{R}^{r\times K}.
$

\item The matrices $\Theta_{c}\in\mathbb{R}^{s\times q_{c}}$, $\Theta_{A}\in\mathbb{R}^{s\times q_{A}}$, and  $\Theta_{H}\in\mathbb{R}^{s\times q_{H}}$ of \cref{eq:Theta-matrices} have full column rank.
\end{enumerate}
If condition 1 holds, then the loss function $\mathscr{L}$ has a global minimizer $\widehat{\mathbf{O}}$ satisfying $\mathscr{L}(\widehat{\mathbf{O}}) = 0$. If conditions 2 and 3 also hold, then that minimizer is unique.

\begin{proof}
Assume condition 1 (no truncation error at the parameter samples). Then the system of ODEs \cref{eq:ode-affine} with operators $\widehat{\mathbf{c}}^{(p)}$, $\widehat{\mathbf{A}}^{(p)}$, and $\widehat{\mathbf{H}}^{(p)}$ derived from \cref{eq:inner_product} holds exactly for all $x\in\Omega$, $t\in[t_0,t_f]$, and $\mu\in\{\mu_{i}\}_{i=1}^{s}$, i.e., it is equivalent to \cref{eq:pde-single}--\cref{eq:pde-bcs-neumann} at the parameter samples. Constructing $\widehat{\mathbf{O}}$ from these operators, we have
$
    \frac{\textup{d}}{\textup{d}t}\widehat{\mathbf{u}}(t;\mu)
    = \mathbf{F}(\widehat{\mathbf{O}};\widehat{\mathbf{u}},t,\theta,\mu)
$
exactly for each $\mu\in\{\mu_{i}\}_{i=1}^{s}$. By construction, $\mathscr{L}(\widehat{\mathbf{O}}) = 0$, which---since $\mathscr{L}$ is non-negative---shows that $\widehat{\mathbf{O}}$ is a global minimizer of $\mathscr{L}$.

To prove uniqueness, assume conditions 2 and 3 and suppose $\widehat{\mathbf{O}}^{(1)}$ and $\widehat{\mathbf{O}}^{(2)}$ both minimize $\mathscr{L}$. By the previous argument, $\mathscr{L}(\widehat{\mathbf{O}}^{(1)}) = \mathscr{L}(\widehat{\mathbf{O}}^{(2)}) = 0$, which implies
\begin{align}
    \label{eq:uniqueness-argument}
    \mathbf{0}_{r}
    &= \mathbf{F}(\widehat{\mathbf{O}}^{(1)};\widehat{\mathbf{u}},\tau_{i,j},\theta,\mu_{i})
     - \mathbf{F}(\widehat{\mathbf{O}}^{(2)};\widehat{\mathbf{u}},\tau_{i,j},\theta,\mu_{i})
    = \mathbf{F}(\widetilde{\mathbf{O}};\widehat{\mathbf{u}},\tau_{i,j},\theta,\mu_{i})
\end{align}
for $i=1,\ldots,s$ and $j=1,\ldots,K$, where $\mathbf{0}_{r}\in\mathbb{R}^{r}$ is a column vector of zeros and
\begin{align*}
    \widetilde{\mathbf{O}}
    &= \widehat{\mathbf{O}}^{(1)} - \widehat{\mathbf{O}}^{(2)}
    = \left[\begin{array}{c|c|c}
        \widetilde{\mathbf{c}}^{(1)}\ \cdots\ \widetilde{\mathbf{c}}^{(q_c)}
        &
        \widetilde{\mathbf{A}}^{(1)}\ \cdots\ \widetilde{\mathbf{A}}^{(q_A)}
        &
        \widetilde{\mathbf{H}}^{(1)}\ \cdots\ \widetilde{\mathbf{H}}^{(q_H)}
    \end{array}\right] \in\mathbb{R}^{r \times q(r)}.
\end{align*}
For fixed $i$, \cref{eq:uniqueness-argument} with $j=1,\ldots,K$ can be written as the linear system
\begin{align*}
    \widetilde{\mathbf{D}}(\mu_{i})
    \left[\begin{array}{c|c|c}
    \sum_{p=1}^{q_{c}}\theta_{c}^{(p)}(\mu_{i})\widetilde{\mathbf{c}}^{(p)}
        &
        \sum_{p=1}^{q_{A}}\theta_{A}^{(p)}(\mu_{i})\widetilde{\mathbf{A}}^{(p)}
        &
        \sum_{p=1}^{q_{H}}\theta_{H}^{(p)}(\mu_{i})\widetilde{\mathbf{H}}^{(p)}
\end{array}\right]^\top
    = \mathbf{0}_{K\times r}, \end{align*}
where $\mathbf{0}_{K\times r}\in\mathbb{R}^{K\times r}$ is the zero matrix. As each $\widetilde{\mathbf{D}}(\mu_i)$ is invertible (condition 2), for $i=1,\ldots,s$ we have
\begin{align*}
    \sum_{p=1}^{q_{c}}\theta_{c}^{(p)}(\mu_{i})\widetilde{\mathbf{c}}^{(p)}
    &= \mathbf{0}_{r}, &
    \sum_{p=1}^{q_{A}}\theta_{A}^{(p)}(\mu_{i})\widetilde{\mathbf{A}}^{(p)}
    &= \mathbf{0}_{r\times r}, &
    \sum_{p=1}^{q_{H}}\theta_{H}^{(p)}(\mu_{i})\widetilde{\mathbf{H}}^{(p)}
    &= \mathbf{0}_{r\times \binom{r+1}{2}}. \end{align*}
Collecting these equations component-wise for $i=1,\ldots,s$ yields the equations
\begin{align*}
    \Theta_{c}\left[\begin{array}{c}
        {[\widetilde{\mathbf{c}}^{(1)}]}_{j}
        \\ \vdots \\
        {[\widetilde{\mathbf{c}}^{(q_c)}]}_{j}
    \end{array}\right]
    =
    \Theta_{A}\left[\begin{array}{c}
        {[\widetilde{\mathbf{A}}^{(1)}]}_{jk}
        \\ \vdots \\
        {[\widetilde{\mathbf{A}}^{(q_A)}]}_{jk}
    \end{array}\right]
    =
    \Theta_{H}\left[\begin{array}{c}
        {[\widetilde{\mathbf{H}}^{(1)}]}_{jk}
        \\ \vdots \\
        {[\widetilde{\mathbf{H}}^{(q_H)}]}_{jk}
    \end{array}\right]
    =
    \mathbf{0}_{s}
\end{align*}
for every index pair $j$,$k$. But $\Theta_c$, $\Theta_{A}$, and $\Theta_{H}$ each have full column rank (condition 3), implying
$
      [\widetilde{\mathbf{c}}^{(p)}]_{j}
    = [\widetilde{\mathbf{A}}^{(p)}]_{jk}
    = [\widetilde{\mathbf{H}}^{(p)}]_{jk}
    = 0
$ for all $j,k,p$. Thus, $\widetilde{\mathbf{O}}$ is the zero matrix, hence $\widehat{\mathbf{O}}^{(1)} = \widehat{\mathbf{O}}^{(2)}$.
\end{proof}
\end{theorem}

\Cref{thm:existence-uniqueness} easily extends to systems of arbitrary polynomial order. Hence, for finite-dimensional PDEs (condition 1) with a polynomial, affine-parametric structure, the operators defined via \cref{eq:inner_product} uniquely minimize the loss function $\mathscr{L}$ defined in \cref{eq:loss-timecontinuous}, provided that the dynamics are sufficiently diverse (as quantified by conditions 2 and 3). As the time step between the data samples defining the pOpInf loss $\mathcal{L}$ of \cref{eq:opinf-loss} decreases, i.e., $t_{j+1} - t_{j} \to 0$ for all $j$, then $\mathcal{L}$ converges to $\mathscr{L}$ (up to a constant), in which case the pOpInf ROM and the intrusive projection-based ROM will be the same. For infinite-dimensional PDEs, condition 1 is realized as the number of basis functions in the approximation increases, i.e., $r\to\infty$. Thus, there are two factors explaining why ROMs learned through pOpInf are not guaranteed to agree with \cref{eq:inner_product}: 1) the truncation error of the representation \cref{eq:truncation-projection}, and 2) the finite sampling of the solution in time. See \cite{Peherstorfer2020reprojection,uy2021probabilistic} for additional work connecting the intrusive and data-driven ROMs.

\section{Affine Operator Inference for Systems of PDEs}
\label{sec:pde_systems}
Systems of partial differential equations with affine polynomial structure admit low-dimensional representations similar to \cref{eq:ode-affine}--\cref{eq:ode-initial-condition} \cite{GW2021learning}. Consider the system of $d$ partial differential equations where each equation can be written as in \cref{eq:pde-single}--\cref{eq:pde-bcs-neumann}:
\begin{subequations}
\begin{align}
    \label{eq:pde-multi}
    \dfrac{\partial u_{\ell}}{\partial t} &= \mathcal{F}_{\ell}(u_{1},\ldots,u_{d};\mu),
    & \ell &= 1,\ldots, d,
\end{align}
where each state variable $u_{\ell}$ is contained in a Hilbert space $\mathcal{V}_{\ell}$ of real-valued functions satisfying appropriate homogeneous boundary conditions. Suppose each spatial differential operator $\mathcal{F}_{\ell}:\mathcal{V}_{1}\times\cdots\times\mathcal{V}_{d}\times\mathcal{P}\to\mathcal{V}_{\ell}^{*}$ is polynomial in the variables $u_1,\ldots,u_d$ and their spatial derivatives,
\begin{align}
    \label{eq:pde-multi-polynomial}
    \begin{aligned}
    \mathcal{F}_{\ell}(u_1,\ldots,u_d;\mu)
    = \mathcal{C}_{\ell}(\mu)
    &+ \sum_{m=1}^{d}\mathcal{A}_{\ell,m}(u_{m};\mu)+ \sum_{m=1}^{d}\sum_{n=m}^{d}\mathcal{H}_{\ell,mn}(u_{m},u_{n};\mu),
\end{aligned}
\end{align}
where $\mathcal{C}_{\ell}:\mathcal{P}\to\mathcal{V}_{\ell}^{*}$, $\mathcal{A}_{\ell,m}:\mathcal{V}_{m}\times\mathcal{P}\to\mathcal{V}_{\ell}^{*}$, $\mathcal{H}_{\ell,mn}:\mathcal{V}_{m}\times\mathcal{V}_{n}\times\mathcal{P}\to\mathcal{V}_{\ell}^{*}$, and each $\mathcal{A}_{\ell,m}$ and $\mathcal{H}_{\ell,mn}$ is linear in each state argument. This is the natural multivariate extension of \cref{eq:pde-polynomial}. We further assume that each operator $\mathcal{C}_{\ell}$, $\mathcal{A}_{\ell,m}$, and $\mathcal{H}_{\ell,mn}$ has an affine-parametric expansion, that is,
\begin{gather}
    \label{eq:affine-expansion-system}
    \begin{gathered}
    \mathcal{C}_{\ell}(\mu)
    = \sum_{p=1}^{q_{c_{\ell}}}\theta_{c_{\ell}}^{(p)}(\mu)\mathcal{C}_{\ell}^{(p)}.
    \qquad
    \mathcal{A}_{\ell,m}(u;\mu)
    = \sum_{p=1}^{q_{A_{\ell,m}}}\theta_{A_{\ell,m}}^{(p)}(\mu)\mathcal{A}_{\ell,m}^{(p)}(u).
    \\
    \mathcal{H}_{\ell,mn}(u,v;\mu)
    = \sum_{p=1}^{q_{H_{\ell,mn}}}\theta_{H_{\ell,mn}}^{(p)}(\mu)\mathcal{H}_{\ell,mn}^{(p)}(u,v).
    \end{gathered}
\end{gather}
\end{subequations}

For each $\ell=1,\ldots, d$, let $\{v_{\ell j}\}_{j=1}^{r_\ell}\subset\mathcal{V}_{\ell}$ be an orthonormal set. We assume a reduced representation $\breve{u}_{\ell}$ of $u_{\ell}$, confined to $\textup{span}(\{v_{\ell_{1}},\ldots,v_{\ell r_{\ell}}\})\subset\mathcal{V}_{\ell}$, given by
\begin{align}
    \label{eq:basis-expansion-multi}
    \breve{u}_{\ell}(x,t;\mu)
    &= \sum_{j=1}^{r_{\ell}} \hat{u}_{\ell j}(t;\mu) v_{\ell j}(x),
    &
    \hat{u}_{\ell j}(t;\mu)
    &= \left\langle v_{\ell j}, u_{\ell}(\cdot,t;\mu) \right\rangle_{\mathcal{V}_{\ell}},
\end{align}
where $\langle\cdot,\cdot\rangle_{\mathcal{V}_{\ell}}$ is the duality pairing of $\mathcal{V}_{\ell}$ with its dual. Inserting $\breve{u}_{\ell}$ into \cref{eq:pde-multi} for each $\ell$ and proceeding as before, we obtain $d$ ODEs that serve as a ROM for \cref{eq:pde-multi}--\cref{eq:affine-expansion-system}:
\begin{subequations}
\begin{align}
    \label{eq:ode-system}
    &\begin{aligned}
    \frac{\textup{d}}{\textup{d} t}\widehat{\mathbf{u}}_{\ell}(t;\mu)
    &= \mathbf{F}_{\ell}(\widehat{\mathbf{O}}_{\ell};\widehat{\mathbf{u}}_{1},\ldots,\widehat{\mathbf{u}}_{d},t,\theta,\mu)
    \\
    &= \left(\sum_{p=1}^{q_{c_{\ell}}}\theta_{c_{\ell}}^{(p)}(\mu)\widehat{\mathbf{c}}_{\ell}^{(p)}\right)
    + \sum_{m=1}^{d}\left(\sum_{p=1}^{q_{A_{\ell,m}}}\theta_{A_{\ell,m}}^{(p)}(\mu)\widehat{\mathbf{A}}_{\ell,m}^{(p)}\right)\widehat{\mathbf{u}}_{m}(t;\mu)
    \\
    &\quad + \sum_{m=1}^{d}\left(\sum_{p=1}^{q_{H_{\ell,mm}}}\theta_{H_{\ell,mm}}^{(p)}(\mu)\widehat{\mathbf{H}}_{\ell,mm}^{(p)}\right)\big(\widehat{\mathbf{u}}_{m}(t;\mu)\,\widehat{\odot}\,\widehat{\mathbf{u}}_{m}(t;\mu)\big)
    \\
    &\quad + \sum_{m=1}^{d}\sum_{n=m+1}^{d}\left(\sum_{p=1}^{q_{H_{\ell,mn}}}\theta_{H_{\ell,mn}}^{(p)}(\mu)\widehat{\mathbf{H}}_{\ell,mn}^{(p)}\right)\big(\widehat{\mathbf{u}}_{m}(t;\mu)\odot\widehat{\mathbf{u}}_{n}(t;\mu)\big),
    \end{aligned}
    \\
    \label{eq:state-vector-system}
    &\widehat{\mathbf{u}}_{\ell}(t;\mu)
    = \left[\begin{array}{ccc}
    \langle v_{\ell 1}, u_{\ell}(\cdot,t;\mu)\rangle & \cdots & \langle v_{\ell,r_{\ell}}, u_{\ell}(\cdot,t;\mu)\rangle
    \end{array}\right]^\top \in \mathbb{R}^{r_{\ell}},
    \\
    \label{eq:operator-matrix-system}
    &\widehat{\mathbf{O}}_{\ell}
    = \left[\begin{array}{ccc|ccc|ccc}
        \widehat{\mathbf{c}}_{\ell}^{(1)}
        & \cdots &
        \widehat{\mathbf{c}}_{\ell}^{(q_{c_\ell})}
        &
        \widehat{\mathbf{A}}_{\ell,1}^{(1)}
        & \cdots &
        \widehat{\mathbf{A}}_{\ell,d}^{(q_{A_{\ell,d}})}
        &
        \widehat{\mathbf{H}}_{\ell,11}^{(1)}
        & \cdots &
        \widehat{\mathbf{H}}_{\ell,dd}^{(q_{H_{\ell,dd}})}
    \end{array}\right],
\end{align}
\end{subequations}
where $\widehat{\mathbf{c}}_{\ell}^{(p)} \in \mathbb{R}^{r_{\ell}}$, $\widehat{\mathbf{A}}_{\ell,m}^{(p)}\in\mathbb{R}^{r_{\ell} \times r_m}$, $\widehat{\mathbf{H}}_{\ell,mm}^{(p)}\in\mathbb{R}^{r_{\ell} \times \binom{r_m+1}{2}}$, and $\widehat{\mathbf{H}}_{\ell,mn}^{(p)}\in\mathbb{R}^{r_{\ell} \times r_m r_n}$ ($m \neq n$). Here $\odot$ denotes the Khatri-Rao product, whereas $\widehat{\odot}$ is the compact Khatri-Rao product that omits redundant terms (see \cref{appendix:kronecker}). Higher-order operators (e.g., cubic terms) can be similarly accounted for.

The low-dimensional system \cref{eq:ode-system} retains the differential and parametric structure of the PDE system \cref{eq:pde-multi}--\cref{eq:pde-multi-polynomial}. The corresponding affine pOpInf problem decouples into $d$ instances of \cref{eq:opinf-standard} which can be solved \emph{independently}:
\begin{align}
    \label{eq:opinf-standard-system}
    &\min_{\widehat{\mathbf{O}}_{\ell}}\left\|
        \mathbf{D}_{\ell}\widehat{\mathbf{O}}_{\ell}^\top - \mathbf{R}_{\ell}^\top
    \right\|_{F}^2,
    &
    \ell &= 1,\ldots, d,
\end{align}
where
\begin{align*}
    \mathbf{R}_{\ell} &= \left[\begin{array}{ccc}
        \dot{\widehat{\mathbf{U}}}_{\ell}(\mu_1)
        & \cdots &
        \dot{\widehat{\mathbf{U}}}_{\ell}(\mu_s)
    \end{array}\right] \in \mathbb{R}^{r_{\ell} \times sK},
    \\
    \dot{\widehat{\mathbf{U}}}_{\ell}(\mu_i) &= \left[\begin{array}{ccc}
        \frac{\textup{d}}{\textup{d}t}\widehat{\mathbf{u}}_{\ell}(t;\mu_i)\Bigr|_{t=t_{1}} & \cdots & \frac{\textup{d}}{\textup{d}t}\widehat{\mathbf{u}}_{\ell}(t;\mu_{i})\Bigr|_{t=t_{K}}
    \end{array}\right] \in \mathbb{R}^{r_{\ell} \times K},
\end{align*}
and where each $\mathbf{D}_{\ell}$ is constructed from the matrices $\widehat{\mathbf{U}}_1(\mu_1),\ldots,\widehat{\mathbf{U}}_d(\mu_s)$, where
\begin{align*}
    \widehat{\mathbf{U}}_{\ell}(\mu_i) &= \left[\begin{array}{ccc}
        \widehat{\mathbf{u}}_{\ell}(t_{1};\mu_i) & \cdots & \widehat{\mathbf{u}}_{\ell}(t_{K};\mu_{i})
    \end{array}\right] \in \mathbb{R}^{r_{\ell} \times K}.
\end{align*}
See \cref{appendix:systems} for the general construction, and note that the number of terms in the PDE dictates the number of terms in the ROM and hence the size of the pOpInf problem. This is best illustrated by example.

\begin{example}[FitzHugh--Nagumo System]
\label{example:fh-n}
Let $\Omega = (0,1)$ and define the parameters $\mu = (\alpha,\beta,\gamma,\varepsilon)\in\mathbb{R}^{4}$. The following system of equations is a simplification of the Hodgkin-Huxley model for activation and deactivation in a spiking neuron \cite{fitzhugh1961impulses,Nagumo1962pulse}:
\begin{subequations}
\begin{align}
    \label{eq:fhn-system}
    \frac{\partial u_{1}}{\partial t}
    &= \varepsilon\frac{\partial^{2} u_{1}}{\partial x^{2}}
        + \frac{1}{\varepsilon}\left(
            - u_{1}^{3} + 1.1 u_{1}^{2} - 0.1 u_{1}
        - u_{2}
        + \alpha\right),
    \\
    \label{eq:fhn-system2}
    \frac{\partial u_{2}}{\partial t}
    &= \beta u_{1} - \gamma u_{2} + \alpha,
\end{align}
with initial conditions $u_{1}(x,t_0) = u_{2}(x,t_0)=0$ and Neumann boundary conditions
\begin{align}
    \label{eq:fhn-bcs}
    \left.\frac{\partial u_{1}}{\partial x}\right|_{x=0}
    &= f(t) := -50000t^{3}e^{-15t},
    &
    \left.\frac{\partial u_{1}}{\partial x}\right|_{x=1}
    &= 0.
\end{align}
\end{subequations}

We write \cref{eq:fhn-system}--\cref{eq:fhn-bcs} in the language of \cref{eq:pde-multi}--\cref{eq:affine-expansion-system} as
\begin{align*}
    \frac{\partial u_{1}}{\partial t}
    &=
    \theta_{c_1}^{(1)}(\mu)\,\mathcal{C}_{1}^{(1)}
    + \theta_{A_{1,1}}^{(1)}(\mu)\,\mathcal{A}_{1,1}^{(1)}(u_1)
    + \theta_{A_{1,1}}^{(2)}(\mu)\,\mathcal{A}_{1,1}^{(2)}(u_1)
    + \theta_{A_{1,2}}^{(1)}(\mu)\,\mathcal{A}_{1,2}^{(2)}(u_2)
    \\
    & \qquad \qquad
    + \theta_{H_{1,11}}^{(1)}(\mu)\,\mathcal{H}_{1,11}^{(1)}(u_1,u_1) + \theta_{G_{1,111}}^{(1)}(\mu)\,\mathcal{G}_{1,111}^{(1)}(u_1,u_1,u_1),
    \\
    \frac{\partial u_{2}}{\partial t}
    &=
    \theta_{c_2}^{(1)}(\mu)\,\mathcal{C}_{2}^{(1)}
    + \theta_{A_{2,1}}^{(1)}(\mu)\,\mathcal{A}_{2,1}^{(1)}(u_1)
    + \theta_{A_{2,2}}^{(1)}(\mu)\,\mathcal{A}_{2,2}^{(1)}(u_2),
\end{align*}
with the operators and the affine coefficient functions given in \cref{table:fhn-terms}. Here, $\mathcal{G}_{1,111}^{(1)}$ represents the cubic nonlinearity $u_{1}^{3}$. This motivates a ROM of the form
\begin{align*}
    \frac{\textup{d}}{\textup{d} t}\widehat{\mathbf{u}}_{1}(t;\mu)
    &=
    \theta_{c_1}^{(1)}(\mu)\,\widehat{\mathbf{c}}_{1}^{(1)}
    + \left(\theta_{A_{1,1}}^{(1)}(\mu)\,\widehat{\mathbf{A}}_{1,1}^{(1)}
    + \theta_{A_{1,1}}^{(2)}(\mu)\,\widehat{\mathbf{A}}_{1,1}^{(2)}\right) \widehat{\mathbf{u}}_1(t;\mu)
    \\ \nonumber & \qquad
    + \theta_{A_{1,2}}^{(1)}(\mu)\,\widehat{\mathbf{A}}_{1,2}^{(1)} \widehat{\mathbf{u}}_2(t;\mu)
    + \theta_{H_{1,11}}^{(1)}(\mu)\,\widehat{\mathbf{H}}_{1,11}^{(1)} \left(\widehat{\mathbf{u}}_1(t;\mu) \,\widehat{\odot}\, \widehat{\mathbf{u}}_1(t;\mu)\right)
    \\ \nonumber & \qquad
    + \theta_{G_{1,111}}^{(1)}(\mu)\,\widehat{\mathbf{G}}_{1,111}^{(1)} \left(\widehat{\mathbf{u}}_1(t;\mu) \,\widehat{\odot}\, \widehat{\mathbf{u}}_1(t;\mu) \,\widehat{\odot}\, \widehat{\mathbf{u}}_1(t;\mu)\right)
    + \theta_{B_1}^{(1)}(\mu)\widehat{\mathbf{B}}_{1}^{(1)}f(t),
    \\
    \frac{\textup{d}}{\textup{d} t}\widehat{\mathbf{u}}_{2}(t;\mu)
    &=
    \theta_{c_2}^{(1)}(\mu)\,\widehat{\mathbf{c}}_{2}^{(1)}
    + \theta_{A_{2,1}}^{(1)}(\mu)\,\widehat{\mathbf{A}}_{2,1}^{(1)} \widehat{\mathbf{u}}_1(t;\mu)
    + \theta_{A_{2,2}}^{(1)}(\mu)\,\widehat{\mathbf{A}}_{2,2}^{(1)} \widehat{\mathbf{u}}_2(t;\mu),
\end{align*}
where the sizes of each discretized operators are listed in \cref{table:fhn-terms}. The operator $\widehat{\mathbf{B}}_1^{(1)}$ accounts for the Neumann boundary condition on $u_1$ in \cref{eq:fhn-bcs}; it has the coefficient function $\theta_{B_1}^{(1)}(\mu)=\varepsilon$. The term $\widehat{\mathbf{G}}_{1,111}^{(1)}$ is the discretization of $\mathcal{G}_{1,111}^{(1)}$.

The corresponding pOpInf problem is \cref{eq:opinf-standard-system} with $d=2$ and
\begin{align*}
    \widehat{\mathbf{O}}_{1}
    &= \left[\begin{array}{ccccccc}
        \widehat{\mathbf{c}}_{1}^{(1)}
        &
        \widehat{\mathbf{B}}_{1}^{(1)}
        &
        \widehat{\mathbf{A}}_{1,1}^{(1)}
        &
        \widehat{\mathbf{A}}_{1,1}^{(2)}
        &
        \widehat{\mathbf{A}}_{1,2}^{(1)}
        &
        \widehat{\mathbf{H}}_{1,11}^{(1)}
        &
        \widehat{\mathbf{G}}_{1,111}^{(1)}
    \end{array}\right] \in \mathbb{R}^{r_1 \times q_1(r_1,r_2)},
    \\
    \mathbf{D}_{1}
    &= \left[\begin{array}{c|c|c|c}
        \mathbf{D}_{c_1} &
        \mathbf{D}_{A_1} &
        \mathbf{D}_{H_1} &
        \mathbf{D}_{G_1}
    \end{array}\right] \in \mathbb{R}^{sK \times q_1(r_1,r_2)},
    \\
    \mathbf{D}_{c_1}
    &= \left[\begin{array}{cc}
        \theta_{c_1}^{(1)}(\mu_1)\mathbf{1}_{K}
        &
        \theta_{B_1}^{(1)}(\mu_1)\mathbf{f}
        \\
        \vdots & \vdots
        \\
        \theta_{c_1}^{(1)}(\mu_s)\mathbf{1}_{K}
        &
        \theta_{B_1}^{(1)}(\mu_s)\mathbf{f}
    \end{array}\right],
    \\
    \mathbf{D}_{A_1}
    &= \left[\begin{array}{ccc}
        \theta_{A_{1,1}}^{(1)}(\mu_1)\widehat{\mathbf{U}}_{1}(\mu_{1})^{\top}
        &
        \theta_{A_{1,1}}^{(2)}(\mu_1)\widehat{\mathbf{U}}_{1}(\mu_{1})^{\top}
        &
        \theta_{A_{1,2}}^{(1)}(\mu_1)\widehat{\mathbf{U}}_{2}(\mu_{1})^{\top}
        \\
        \vdots & \vdots & \vdots
        \\
        \theta_{A_{1,1}}^{(1)}(\mu_s)\widehat{\mathbf{U}}_{1}(\mu_{s})^{\top}
        &
        \theta_{A_{1,1}}^{(2)}(\mu_s)\widehat{\mathbf{U}}_{1}(\mu_{s})^{\top}
        &
        \theta_{A_{1,2}}^{(1)}(\mu_s)\widehat{\mathbf{U}}_{2}(\mu_{s})^{\top}
    \end{array}\right],
    \\
    \mathbf{D}_{H_1}
    &= \left[\begin{array}{c}
        \theta_{H_{1,11}}^{(1)}(\mu_1)\left(\widehat{\mathbf{U}}_{1}(\mu_{1})\,\widehat{\odot}\,\widehat{\mathbf{U}}_{1}(\mu_{1})\right)^{\top}
        \\
        \vdots
        \\
        \theta_{H_{1,11}}^{(1)}(\mu_s)\left(\widehat{\mathbf{U}}_{1}(\mu_{s})\,\widehat{\odot}\,\widehat{\mathbf{U}}_{1}(\mu_{s})\right)^{\top}
    \end{array}\right],
    \\
    \mathbf{D}_{G_1}
    &= \left[\begin{array}{c}
        \theta_{G_{1,111}}^{(1)}(\mu_1)\left(\widehat{\mathbf{U}}_{1}(\mu_{1})\,\widehat{\odot}\,\widehat{\mathbf{U}}_{1}(\mu_{1})\,\widehat{\odot}\,\widehat{\mathbf{U}}_{1}(\mu_{1})\right)^{\top}
        \\
        \vdots
        \\
        \theta_{G_{1,111}}^{(1)}(\mu_s)\left(\widehat{\mathbf{U}}_{1}(\mu_{s})\,\widehat{\odot}\,\widehat{\mathbf{U}}_{1}(\mu_{s})\,\widehat{\odot}\,\widehat{\mathbf{U}}_{1}(\mu_{s})\right)^{\top}
    \end{array}\right],
    \\
    \widehat{\mathbf{O}}_{2}
    &= \left[\begin{array}{ccc}
        \widehat{\mathbf{c}}_{2}^{(1)}
        &
        \widehat{\mathbf{A}}_{2,1}^{(1)}
        &
        \widehat{\mathbf{A}}_{2,2}^{(1)}
    \end{array}\right] \in \mathbb{R}^{r_{2} \times q_2(r_1,r_2)},
    \\
    \mathbf{D}_{2}
    &= \left[\begin{array}{ccc}
        \theta_{c_2}^{(1)}(\mu_1)\mathbf{1}_{K}
        &
        \theta_{A_{2,1}}^{(1)}(\mu_1)\widehat{\mathbf{U}}_{1}(\mu_{1})^\top
        &
        \theta_{A_{2,2}}^{(1)}(\mu_1)\widehat{\mathbf{U}}_{2}(\mu_{1})^\top
        \\
        \vdots & \vdots & \vdots
        \\
        \theta_{c_2}^{(1)}(\mu_s)\mathbf{1}_{K}
        &
        \theta_{A_{2,1}}^{(1)}(\mu_s)\widehat{\mathbf{U}}_{1}(\mu_{s})^\top
        &
        \theta_{A_{2,2}}^{(1)}(\mu_s)\widehat{\mathbf{U}}_{2}(\mu_{s})^\top
    \end{array}\right] \in \mathbb{R}^{sK \times q_2(r_1,r_2)},
\end{align*}
where $\mathbf{f} = [f(t_1)~\cdots~f(t_K)]^\top\in\mathbb{R}^{K}$,
$q_{1}(r_1,r_2) = 2 + 2r_{1} + r_{2} + \binom{r_{1}+1}{2} + \binom{r_{1}+2}{3}$, and $q_{2}(r_1,r_2) = 1 + r_{1} + r_{2}$. The data matrix $\mathbf{D}_1$ and the operator matrix $\widehat{\mathbf{O}}_1$ have been modified from \cref{eq:ode-system}--\cref{eq:operator-matrix-system} to account for the cubic term present \cref{eq:fhn-system}.
\end{example}

\begin{remark}
\label{remark:FHN-DEIM}
Even though since \cite{CS2010deim}, the FitzHugh--Nagumo system has been widely used as a benchmark problem for the development of general nonlinear model reduction methods, a fully cubic intrusive model can be directly derived for this system (i.e., the nonlinear terms in the system are point-wise local and exactly quadratic and cubic), circumventing the need for the second layer of approximation introduced through hyper-reduction.
\end{remark}

\begingroup
\renewcommand{\arraystretch}{1.5}
\begin{table}[ht]
    \centering
    {\footnotesize
    \caption{The operators in the continuous setting, the associated affine coefficient functions, and the size of the reduced operators in discretized setting for the FitzHugh--Nagumo system of \cref{example:fh-n}.}
    \label{table:fhn-terms}
    \begin{center}
    \begin{tabular}{cllll}
        \toprule
        & & continuous term & affine coefficient & discretized operator\\
        \midrule
        \multirow{7}{*}{\rotatebox[origin=c]{90}{Eq.~\cref{eq:fhn-system}}} & constant & $\mathcal{C}_{1}^{(1)} =1$ & $\theta_{c_1}^{(1)}(\mu)= {\alpha}/{\varepsilon}$ & $\widehat{\mathbf{c}}_{1}^{(1)}\in \mathbb{R}^{r_1}$\\
        & input & $\frac{\partial u_1}{\partial x}\big|_{x=0} = f(t)$ & $\theta_{B_1}^{(1)}(\mu)=\varepsilon$ & $\widehat{\mathbf{B}}_{1}^{(1)}\in \mathbb{R}^{r_1}$\\
        & \multirow{3}{*}{linear} & $\mathcal{A}_{1,1}^{(1)}(u) = \frac{\partial^{2} u}{\partial x^{2}}$ & $\theta_{A_{1,1}}^{(1)}(\mu) = \varepsilon$ & $\widehat{\mathbf{A}}_{1,1}^{(1)}\in \mathbb{R}^{r_1\times r_1}$\\
        & & $\mathcal{A}_{1,1}^{(2)}(u) =u$ & $\theta_{A_{1,1}}^{(2)}(\mu) = -{0.1}/{\varepsilon}$ & $\widehat{\mathbf{A}}_{1,1}^{(2)}\in \mathbb{R}^{r_1\times r_1}$\\
        & & $\mathcal{A}_{1,2}^{(1)}(u) =u$ & $\theta_{A_{1,2}}^{(1)}(\mu) = -{1}/{\varepsilon}$ & $\widehat{\mathbf{A}}_{1,2}^{(1)}\in \mathbb{R}^{r_1\times r_2}$\\
        & quadratic & $\mathcal{H}_{1,11}^{(1)}(u,v) = uv$ & $\theta_{H_{1,11}}^{(1)}(\mu) = {1.1}/{\varepsilon}$ & $\widehat{\mathbf{H}}_{1,11}^{(1)}\in \mathbb{R}^{r_1\times \binom{r_1+1}{2}}$\\
        & cubic & $\mathcal{G}_{1,111}^{(1)}(u,v,w) = uvw$ & $\theta_{G_{1,111}}^{(1)}(\mu) = -{1}/{\varepsilon}$ & $\widehat{\mathbf{G}}_{1,111}^{(1)}\in \mathbb{R}^{r_1\times \binom{r_1+2}{3}}$\\
        \midrule
        \multirow{3}{*}{\rotatebox[origin=c]{90}{Eq.~\cref{eq:fhn-system2}}} & constant & $\mathcal{C}_{2}^{(1)} = 1$ & $\theta_{c_2}^{(1)}(\mu)= \alpha$ & $\widehat{\mathbf{c}}_{2}^{(1)}\in \mathbb{R}^{r_2}$\\
        & \multirow{2}{*}{linear} & $\mathcal{A}_{2,1}^{(1)}(u) =u$ & $\theta_{A_{2,1}}^{(1)}(\mu) = \beta$ & $\widehat{\mathbf{A}}_{2,1}^{(1)}\in \mathbb{R}^{r_2\times r_1}$\\
        & & $\mathcal{A}_{2,2}^{(1)}(u) =u$ & $\theta_{A_{2,2}}^{(1)}(\mu) = -\gamma$ & $\widehat{\mathbf{A}}_{2,2}^{(1)}\in \mathbb{R}^{r_2\times r_2}$\\
        \bottomrule
    \end{tabular}
    \end{center}
    }
\end{table}
\endgroup

\begin{remark}
\label{remark:monolithic}
An alternative approach to constructing a ROM for \cref{eq:pde-multi}--\cref{eq:affine-expansion-system} is to consider the variables $u_{1},\ldots,u_{d}$ as members of a product Hilbert space \cite{qian2022opinfpde}. Let $\{\vec{v}_{j}\}_{j=1}^{r} \subset \mathcal{V}_{\times} = \mathcal{V}_{1}\times\cdots\times\mathcal{V}_{d}$ be orthonormal with respect to the natural inner product
\begin{align*}
    \left\langle\vec{u},\vec{w}\right\rangle_{\mathcal{V}_{\times}}
    = \sum_{\ell=1}^{d}\left\langle u_{\ell}, w_{\ell}\right\rangle_{\mathcal{V}_{\ell}},
\end{align*}
where $\vec{u} = (u_{1},\ldots,u_{d})$ and $\vec{w} = (w_{1},\ldots,w_{d})$. In this case, Galerkin projection yields a low-dimensional system with the structure of \cref{eq:ode-affine}--\cref{eq:ode-initial-condition} where for each operator, the parametric coefficient functions are the union of the parametric coefficient functions for the operators of the same order from \cref{eq:affine-expansion-system}. That is,
\begin{gather*}
    \{\theta_{c}^{(p)}\}_{p=1}^{q_{c}}
    = \bigcup_{\ell=1}^{d}\{\theta_{c_{\ell}}^{(p)}\}_{p=1}^{q_{c_{\ell}}},
    \ \ \ \
    \{\theta_{A}^{(p)}\}_{p=1}^{q_{A}}
    = \bigcup_{\ell=1}^{d}\{\theta_{A_{\ell}}^{(p)}\}_{p=1}^{q_{A_{\ell}}},
    \ \ \ \
    \{\theta_{H}^{(p)}\}_{p=1}^{q_{H}}
    = \bigcup_{\ell=1}^{d}\{\theta_{H_{\ell}}^{(p)}\}_{p=1}^{q_{H_{\ell}}}.
\end{gather*}
See \cite{YGBK2021} for details and an application to the shallow water equations. We call this approach \emph{monolithic} because there is a single set of basis functions for all variables jointly, not individual basis functions corresponding to each variable as described in \cref{eq:basis-expansion-multi}. The parametric dependencies in the corresponding ROM are determined by the governing equations, but the full system-level structure is not preserved.
\end{remark}

\begin{example}[Monolithic FitzHugh--Nagumo System]
With a monolithic basis as described in \cref{remark:monolithic}, the FitzHugh--Nagumo system \cref{eq:fhn-system}--\cref{eq:fhn-bcs} motivates a ROM with the form
\begin{align*}
    \frac{\textup{d}}{\textup{d}t}\widehat{\mathbf{u}}(t;\mu)
    &= \frac{\alpha}{\varepsilon}\widehat{\mathbf{c}}^{(1)} + \alpha\widehat{\mathbf{c}}^{(2)}
    + \left(
        \varepsilon\widehat{\mathbf{A}}^{(1)}
        -\frac{1}{\varepsilon}\widehat{\mathbf{A}}^{(2)}
        + \beta\widehat{\mathbf{A}}^{(3)}
        - \gamma\widehat{\mathbf{A}}^{(4)}
    \right)\widehat{\mathbf{u}}(t;\mu)
    \\&\:
    + \frac{1}{\varepsilon}\widehat{\mathbf{H}}^{(1)}\big(\widehat{\mathbf{u}}(t;\mu)\,\widehat{\odot}\,\widehat{\mathbf{u}}(t;\mu)\big)
    -\frac{1}{\varepsilon}\widehat{\mathbf{G}}^{(1)}\big(\widehat{\mathbf{u}}(t;\mu)\,\widehat{\odot}\,\widehat{\mathbf{u}}(t;\mu)\,\widehat{\odot}\,\widehat{\mathbf{u}}(t;\mu)\big)
    + \varepsilon\widehat{\mathbf{B}}^{(1)}f(t).
\end{align*}
Contrast this with the ROM prescribed in \cref{example:fh-n}, which inherits the entire structure of the PDE system \cref{eq:fhn-system}--\cref{eq:fhn-system2}. In particular, the reduced-order operators in \cref{example:fh-n} correspond directly to individual PDE operators (see \cref{table:fhn-terms}).
\end{example}

\section{Computational Procedure}
\label{sec:computation}
Solving the pOpInf problem \cref{eq:opinf-standard} requires samples of the solution $u(x,t;\mu)$ and its time derivative at times $\{t_{j}\}_{j=1}^{K}$ for each selected parameter value $\{\mu_{i}\}_{i=1}^{s}$. The quality of the resulting ROM depends on how well the orthonormal basis functions $\{v_{j}\}_{j=1}^{r}$ represent the solution at each parameter value and throughout the spatial and temporal domains. We therefore adopt the widely used proper orthogonal decomposition (POD) \cite{algazi1969optimality,Berkooz1993,lumley,sirovich1987turbulence}, defined by the set of orthonormal functions that minimize the mean squared projection error of the sample data, i.e., solving the problem
\begin{align*}
    &\min_{v_{1},\ldots,v_{r}\in\mathcal{V}}
        \sum_{i=1}^{s}\sum_{j=1}^{K}\left\|
            u(\cdot,t_{j};\mu_{i})
            - \sum_{\ell=1}^{r}\big\langle
                v_{\ell},u(\cdot,t_j;\mu_i)
                \big\rangle v_{\ell}
        \right\|_{\mathcal{V}}^2
    &
    &\textrm{subject to}
    &
    &\langle v_{i}, v_{j} \rangle = \delta_{ij},
\end{align*}
where $\|v\|_{\mathcal{V}} = \sqrt{\langle v,v\rangle}$ is the natural norm on $\mathcal{V}$. This data-driven choice of basis optimally represents the solution at the sampled parameter values, although the resulting ROM does not share such guarantees \cite{BGW2015pmorSurvery}.

The conditioning of the linear least-squares problem \cref{eq:opinf-standard} depends on the data matrix $\mathbf{D}$. If the parameter samples are chosen so that the pitfalls described in \cref{thm:rank-deficiencies} are avoided, then the condition number of $\mathbf{D}$ depends on the nature of the solution at the quadrature points $\{t_{j}\}_{j=1}^{K}$. To improve the conditioning, we introduce a Tikhonov regularization \cite{Tikhonov1977regularization} so that \cref{eq:opinf-standard} becomes
\begin{align}
    \label{eq:opinf-reg}
    &\min_{\widehat{\mathbf{O}}}\left\|
        \mathbf{D}\widehat{\mathbf{O}}^\top - \mathbf{R}^\top
    \right\|_{F}^2
    + \left\|\boldsymbol{\Lambda}\widehat{\mathbf{O}}^\top\right\|_F^2,
    &
    & \boldsymbol{\Lambda} \in\mathbb{R}^{q(r)\times q(r)}.
\end{align}
The solution to this regularized problem satisfies the modified normal equations,
\begin{align}
    \label{eq:normal-equations}
    \left(
        \mathbf{D}^\top\mathbf{D}
        + \boldsymbol{\Lambda}^\top\boldsymbol{\Lambda}
    \right)
    \widehat{\mathbf{O}}^\top
    = \mathbf{D}^{\top}\mathbf{R}^\top.
\end{align}
The regularizer $\boldsymbol{\Lambda}$ can be parameterized in a number of ways \cite{MHW2021regOpInfCombustion,qian2022opinfpde}. One choice that provides flexibility without introducing a large number of hyperparameters is the diagonal matrix $\boldsymbol{\Lambda} = \boldsymbol{\Lambda}(\lambda_1,\lambda_2)$ defined such that
\begin{align}
    \label{eq:regularization-explicit}
    \left\|
        \boldsymbol{\Lambda}(\lambda_1,\lambda_2)
        \widehat{\mathbf{O}}^\top
    \right\|_F^2
    = \lambda_1^{2}\left(
        \sum_{p=1}^{q_{c}}\left\|\widehat{\mathbf{c}}^{(p)}\right\|_{2}^{2} +
        \sum_{p=1}^{q_{A}}\left\|\widehat{\mathbf{A}}^{(p)}\right\|_{F}^{2}
        \right)
    + \lambda_2^{2}
        \sum_{p=1}^{q_{H}}\left\|\widehat{\mathbf{H}}^{(p)}\right\|_{F}^{2}.
\end{align}
This regularization structure groups the operators defining the ROM according to their polynomial order and drives $\widehat{\mathbf{O}}$ toward the zero matrix as $\lambda_{1},\lambda_{2}\to \infty$. Therefore, the regularization drives the resulting ROM toward the globally stable zero system $\frac{\textup{d}}{\textup{d}t}\widehat{\mathbf{u}}(t;\mu) = \mathbf{0}$. As before, the extension of \cref{eq:opinf-reg} to a system of PDEs is straightforward where each PDE is written as an independent pOpInf problem (see \cref{eq:opinf-standard-system}).

We choose the regularization hyperparameters $\lambda_1,\lambda_2 \ge 0$ to minimize the mean squared training error
\begin{align*}
\frac{1}{s}\sum_{i=1}^{s}\sum_{j=1}^{K}\left\|
        \widehat{\mathbf{u}}(t_{j};\mu_{i}) -
        \widetilde{\mathbf{u}}(t_{j};\mu_{i})
    \right\|_{2}^{2},
\end{align*}
where $\widehat{\mathbf{u}}(t_{j};\mu_{i})$ is the training data and $\widetilde{\mathbf{u}}(t_{j};\mu_{i})$ is the result of integrating the ODE $\frac{\textrm{d}}{\textrm{d}t}\widetilde{\mathbf{u}}(t) = \mathbf{F}(\widehat{\mathbf{O}};\widetilde{\mathbf{u}},t,\mu)$ defined by the solution $\widehat{\mathbf{O}}$ of \cref{eq:opinf-reg} with regularization hyperparameters $\lambda_1$ and $\lambda_2$. This is an optimization problem in the principal quadrant of $\mathbb{R}^{2}$, which we carry out with a sparse grid search followed by a derivative-free search method \cite{nelder1965simplex}. The approach guarantees that the selected hyperparameters result in a ROM that is stable for all training parameter values $\mu_{1},\ldots,\mu_{s}$. To further promote ROM stability throughout the parameter domain $\mathcal{P}$, we may constrain the hyperparameter selection problem by introducing a set of $\bar{s}$ parameters $\{\bar{\mu}_{i}\}_{i=1}^{\bar{s}} \subset \mathcal{P}$ at which we demand stability from the learned ROM. Specifically, we disqualify hyperparameter pairings $(\lambda_{1},\lambda_{2})$ whenever the time integrator for the resulting ROM diverges for any $\mu \in \{\bar{\mu}_{i}\}_{i=1}^{\bar{s}}$. This approach is similar to the hyperparameter selection strategy of \cite{MHW2021regOpInfCombustion} for non-parametric problems wherein the objective is to predict beyond the temporal domain for which data are available; our strategy focuses on the prediction in the parametric space but could also be adapted to address prediction in time.

We now summarize the computational procedure for solving \cref{eq:opinf-standard}: \hypertarget{steps:step1}{1)}~select parameter values $\{\mu_{i}\}_{i=1}^{s}$ such that the associated $\Theta_{c},\ldots,\Theta_{G}$ have full column rank; \hypertarget{steps:step2}{2)}~sample the PDE solution $u(x,t,\mu)$ and its time derivative for $t\in\{t_{j}\}_{j=1}^{K}$ for each $\mu\in\{\mu_{i}\}_{i=1}^{s}$; \hypertarget{steps:step3}{3)}~compute the POD basis associated with the sampled solution data; \hypertarget{steps:step4}{4)}~use the POD basis to project the solution data, obtaining $\widehat{\mathbf{U}}(\mu_i)$ and $\dot{\widehat{\mathbf{U}}}(\mu_{i})$; \hypertarget{steps:step5}{5)}~form the data matrix $\mathbf{D}$ and the time derivative matrix $\mathbf{R}$; \hypertarget{steps:step6}{6)}~choose optimal regularization hyperparameters and solve \cref{eq:opinf-reg} with these hyperparameters. In the computational setting, we sample the solution in \hyperlink{steps:step2}{step 2} by obtaining approximate discretized \emph{solution snapshots} via a high-fidelity solver. For example, let $\{\mathbf{x}_{\ell}\}_{\ell=1}^{N}\subset\Omega$ be a discretization of $\Omega$, and define
\begin{align*}
    \mathbf{U}(\mu_{i})
    &= \left[\begin{array}{ccc}
        u(\mathbf{x}_{1},t_{1};\mu_i) & \cdots & u(\mathbf{x}_{1},t_{K};\mu_i)
        \\ \vdots & & \vdots \\
        u(\mathbf{x}_{N},t_{1};\mu_i) & \cdots & u(\mathbf{x}_{N},t_{K};\mu_i)
    \end{array}\right]\in\mathbb{R}^{N\times K},
\end{align*}
the \emph{snapshot matrix} for parameter $\mu_{i}$. The rank-$r$ POD basis of \hyperlink{steps:step3}{step 3} is comprised of the first $r$ left singular vectors of the concatenated snapshot matrices, that is,
\begin{align}
    \label{eq:pod-from-svd}
    \boldsymbol{\Phi}\boldsymbol{\Sigma}\boldsymbol{\Psi}^\top
    &= \left[\begin{array}{ccc}
        & & \\
        \mathbf{U}(\mu_{1}) & \cdots & \mathbf{U}(\mu_{s})
        \\ & &
    \end{array}\right]\in\mathbb{R}^{N\times sK},
    &
    \mathbf{V} &= \boldsymbol{\Phi}_{:,1:r}\in\mathbb{R}^{N\times r},
\end{align}
where $\boldsymbol{\Phi}\boldsymbol{\Sigma}\boldsymbol{\Psi}^\top$ is the singular value decomposition (SVD). With this notation, the projection of \hyperlink{steps:step4}{step 4} is given by $\widehat{\mathbf{U}}(\mu_i) = \mathbf{V}^\top\mathbf{U}(\mu_i)$, $i=1,\ldots,s$. If the time derivatives of $u$ are not provided by the high-fidelity solver, they may be estimated as finite differences of the solution snapshots. The time integration error of the high-fidelity model, as well as the approximation error accompanying finite differences for the time derivatives, are additional motivations for utilizing the regularization strategy described previously. \Cref{alg:OpInf-reg-multi} fully details the procedure. The algorithm is presented for the general case of a system of $d$ partial differential equations, but it may be simplified to a case of a single PDE by setting $d=1$.

\begin{algorithm}
\begin{algorithmic}[1]
\Procedure{pOpInf}{\newline\phantom{---} training parameter values $\mu_{1},\ldots,\mu_{s}\in\mathcal{P}$,
\newline\phantom{---} training snapshots $\mathbf{U}_{1}(\mu_{i}),\ldots,\mathbf{U}_{d}(\mu_{i})\in\mathbb{R}^{N\times K}$ for $i=1,\ldots,s$,
    \newline\phantom{---} affine coefficient functions $\theta = \{\theta_{c_1}^{(1)},\ldots,\theta_{H_{d,dd}}^{(q_{H_{d,dd}})}\}:\mathcal{P}\to\mathbb{R}$,
    \newline\phantom{---} reduced dimensions $r_{1},\ldots,r_{d}\in\mathbb{N}$,
    \newline\phantom{---} stability parameter values $\bar{\mu}_{1},\ldots,\bar{\mu}_{\bar{s}}\in\mathcal{P}$ (optional, else $\bar{s} = 0$)
    \newline}

    \LineComment{Project training data to low-dimensional subspaces.}
    \For{$\ell = 1, \ldots, d$}
        \State $\mathbf{V}_{\ell} \gets\ $\texttt{pod}$\left([\mathbf{U}_{\ell}(\mu_1)~\cdots~\mathbf{U}_{\ell}(\mu_s)],r_{\ell}\right)$
            \Comment{Rank-$r_\ell$ POD basis.}
        \For{$i=1,\ldots,s$}
            \State $\widehat{\mathbf{U}}_{\ell}(\mu_i) \gets \mathbf{V}_{\ell}^\top\mathbf{U}_{\ell}(\mu_i)$
            \Comment{Projected solution data.}
            \State $\dot{\widehat{\mathbf{U}}}_{\ell}(\mu_i) \gets \frac{\textup{d}}{\textup{d}t}\widehat{\mathbf{U}}_{\ell}(\mu_i)$
            \Comment{Projected time derivatives.}
        \EndFor
    \EndFor

    \LineComment{Construct pOpInf matrices.}
    \For{$\ell = 1,\ldots,d$}
        \State $\mathbf{D}_{\ell} \gets$ build the $\ell$th data matrix from $ \widehat{\mathbf{U}}_{1}(\mu_{1}),\ldots,\widehat{\mathbf{U}}_{d}(\mu_{s}),\theta,\mu_{1},\ldots,\mu_{s}$
        \State $\mathbf{R}_{\ell} \gets [\dot{\widehat{\mathbf{U}}}_{\ell}(\mu_1)~\cdots~\dot{\widehat{\mathbf{U}}}_{\ell}(\mu_s)]$
    \EndFor

    \LineComment{Compute pOpInf solution with optimal hyperparameters.}
    \State $\lambda_1^{*},\lambda_2^{*} \gets$ \textrm{argmin}
    \textproc{TrainingError}$(\lambda_1,\lambda_2)$ \State \textbf{return} $\textproc{RegOpInf}(\lambda_{1}^{*},\lambda_{2}^{*})$
\EndProcedure

\vspace{6pt}
\setcounter{ALG@line}{0}
\Procedure{TrainingError} {$\lambda_1,\lambda_2$}
\State $\widehat{\mathbf{O}}_{1},\ldots,\widehat{\mathbf{O}}_{d} \gets \textproc{RegOpInf}(\lambda_{1},\lambda_{2})$
    \If{$\bar{s} > 0$}
    \For{$i = 1, \ldots, \bar{s}$}
    \Comment{Check behavior at stability parameters.}
        \If{integrating \cref{eq:ode-system}, $\ell=1,\ldots,d$, $\mu = \bar{\mu}_{i}$, over $[t_{0},t_{f}]$ diverges}
            \State \textbf{return} $\infty$
        \EndIf
    \EndFor
    \EndIf
    \For{$i = 1, \ldots, s$}
    \Comment{Calculate error at training parameters.}
        \State $\widetilde{\mathbf{U}}_{1}(\mu_{i}),\ldots,\widetilde{\mathbf{U}}_{d}(\mu_{i}) \gets\ $ integrate \cref{eq:ode-system}, $\ell=1,\dots,d$, $\mu = \mu_{i}$, over $[t_0,t_f]$
    \EndFor
    \State \textbf{return} $\frac{1}{sd}\sum_{\ell=1}^{d}\sum_{i=1}^{s}\|\widehat{\mathbf{U}}_{\ell}(\mu_{i}) - \widetilde{\mathbf{U}}_{\ell}(\mu_{i})\|_F^2$
\EndProcedure

\vspace{6pt}
\setcounter{ALG@line}{0}
\Procedure{RegOpInf}{$\lambda_1,\lambda_2$}
    \State $\boldsymbol{\Lambda}^2 \gets \boldsymbol{\Lambda}(\lambda_1,\lambda_2)^\top \boldsymbol{\Lambda}(\lambda_1,\lambda_2)$
    \Comment{Construct the regularizer.}
    \For{$\ell = 1, \ldots, d$}
        \State $\widehat{\mathbf{O}}_{\ell}^\top \gets \left(\mathbf{D}_{\ell}^\top\mathbf{D}_{\ell} + \boldsymbol{\Lambda}^2\right)^{-1}\mathbf{D}_{\ell}^\top\mathbf{R}_{\ell}^\top$
        \Comment{Solve the $\ell$th pOpInf problem.}
    \EndFor
    \State \textbf{return} $\widehat{\mathbf{O}}_{1},\ldots,\widehat{\mathbf{O}}_{d}$
\EndProcedure
\end{algorithmic}
\caption{Regularized \textbf{p}arametric \textbf{Op}erator \textbf{Inf}erence for systems of PDEs}
\label{alg:OpInf-reg-multi}
\end{algorithm}

\begin{remark}
There are several alternatives for addressing poor conditioning in \cref{eq:opinf-standard}. Notably, \cite{YGBK2021} uses a truncated QR decomposition to solve a column subset selection problem, in which the size of the truncation is an integer hyperparameter that is typically selected via the $L$-curve criteria \cite{hansen1992lcurve}. In some settings, building physical constraints into \cref{eq:opinf-standard} also acts as an implicit regularization (see, e.g., \cite{sharma2022hamiltonian}). We choose the Tikhonov regularization defined in \cref{eq:regularization-explicit} because 1) it features a continuous hyperparameter search space, enabling a precise selection; and 2) the structure of the regularizer can be tailored to specific model structure, as we will see in \cref{subsec:numerical_fhn}.
\end{remark}

\section{Numerical Examples}
\label{sec:numerics}
We now present numerical results for the heat equation and FitzHugh--Nagumo system introduced in \Cref{sec:methodology} and \Cref{sec:pde_systems}, respectively. In each of these examples we learn ROMs using \cref{alg:OpInf-reg-multi} and compare their performance to classical ROMs obtained through intrusive projection \cite{BGW2015pmorSurvery}. However, our pOpInf methodology applies to situations in which intrusive projection is infeasible due to, e.g., variable transformations to induce polynomial structure \cite{QKPW2020liftAndLearn}. The code for these experiments can be found at \url{https://github.com/Willcox-Research-Group/affine-parametric-opinf}.

\subsection{Heat Equation}
\label{subsec:numerical_heat}
\begin{figure}
    \centering
    \includegraphics[width=\textwidth]{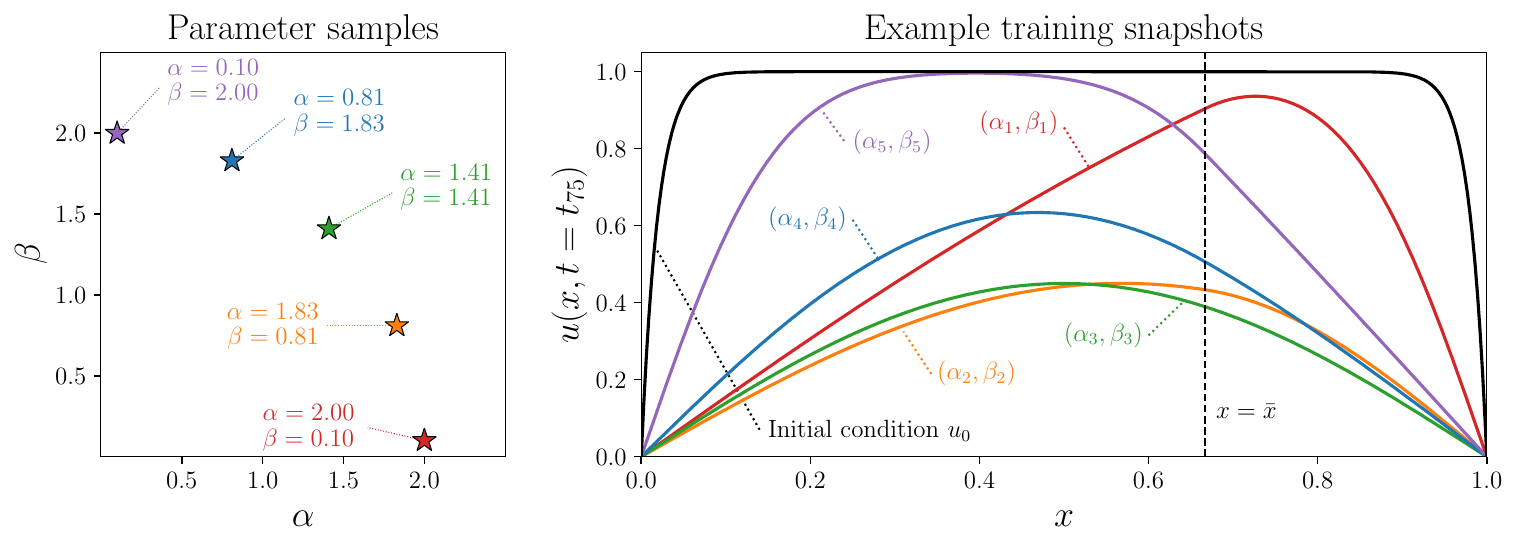}
    \vspace{-0.75cm}
    \caption{Experimental parameter samples (left) and associated snapshots at intermediate time $t = t_{75} = 0.075$ (right) for the heat equation problem \cref{eq:heat-pde}--\cref{eq:heat-bcs}. The vertical line $x = \bar{x}$ marks the point in the domain where the diffusion constant switches between $\alpha$ and $\beta$.}
    \label{fig:heat:training-samples}
\end{figure}

\begin{figure}
    \centering
    \includegraphics[width=\textwidth]{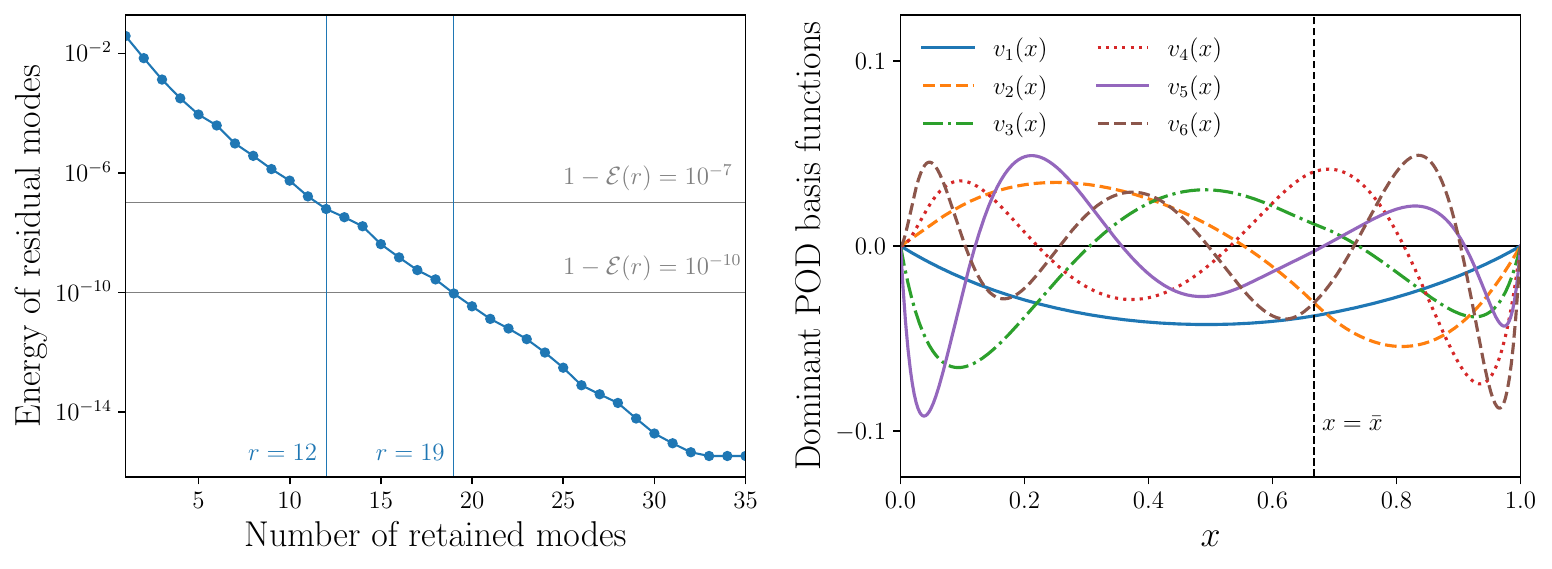}
    \vspace{-0.75cm}
    \caption{Residual energy decay (left) and the six dominant POD basis functions (right) for the snapshot set generated at the parameter samples in \cref{fig:heat:training-samples}. For $r \ge 12$, we have $1 - \mathcal{E}(r) < 10^{-7}$; if $r \ge 19$, then $1 - \mathcal{E}(r) < 10^{-10}$.}
    \label{fig:heat:basis}
\end{figure}

We return to the heat equation of \cref{example:heat-1} and \ref{example:heat-2}, setting $\bar{x} = 2/3$ and
$
    u_0(x;\mu)
    = 1 - \left(1 - x\right)^{50} - x^{50}.
$
For each of the $s = 5$ parameter samples shown in \cref{fig:heat:training-samples}, we generate solutions by discretizing the spatial domain $\Omega = (0,1)$ with a uniform grid of $N = 1000$ points and approximating the spatial derivative with second-order central finite differences. The resulting semi-discrete ODE, called the full-order model, is integrated in time with the first-order implicit Euler scheme on $K = 1500$ uniformly spaced time steps in $[t_0,t_f] = [0,1.5]$. \Cref{fig:heat:training-samples} also shows example snapshots for each parameter sample, demonstrating that variation in the parameters $\mu = (\alpha,\beta)$ determines the diffusion dynamics. To select the number of modes in the POD basis, define the cumulative energy
\begin{align*}
    \mathcal{E}(r)
    = \sum_{j=1}^{r}\sigma_{j}^{2} \bigg/\sum_{j=1}^{N}\sigma_{j}^{2},
\end{align*}
where $\sigma_{j}$ is the $j$th singular value in the POD factorization \cref{eq:pod-from-svd}. Note that $\mathcal{E}(r)$ is a nondecreasing function of $r$. We select $r$ to be the smallest integer such that the residual energy $1 - \mathcal{E}(r)$, or the energy in the non-retained modes, lies below a fixed threshold $\epsilon > 0$. For this problem, setting $\epsilon = 10^{-7}$ results in $r = 12$, while setting $\epsilon = 10^{-10}$ results in $r = 19$ (see \cref{fig:heat:basis}).

\begin{figure}
    \centering
    \includegraphics[width=\textwidth]{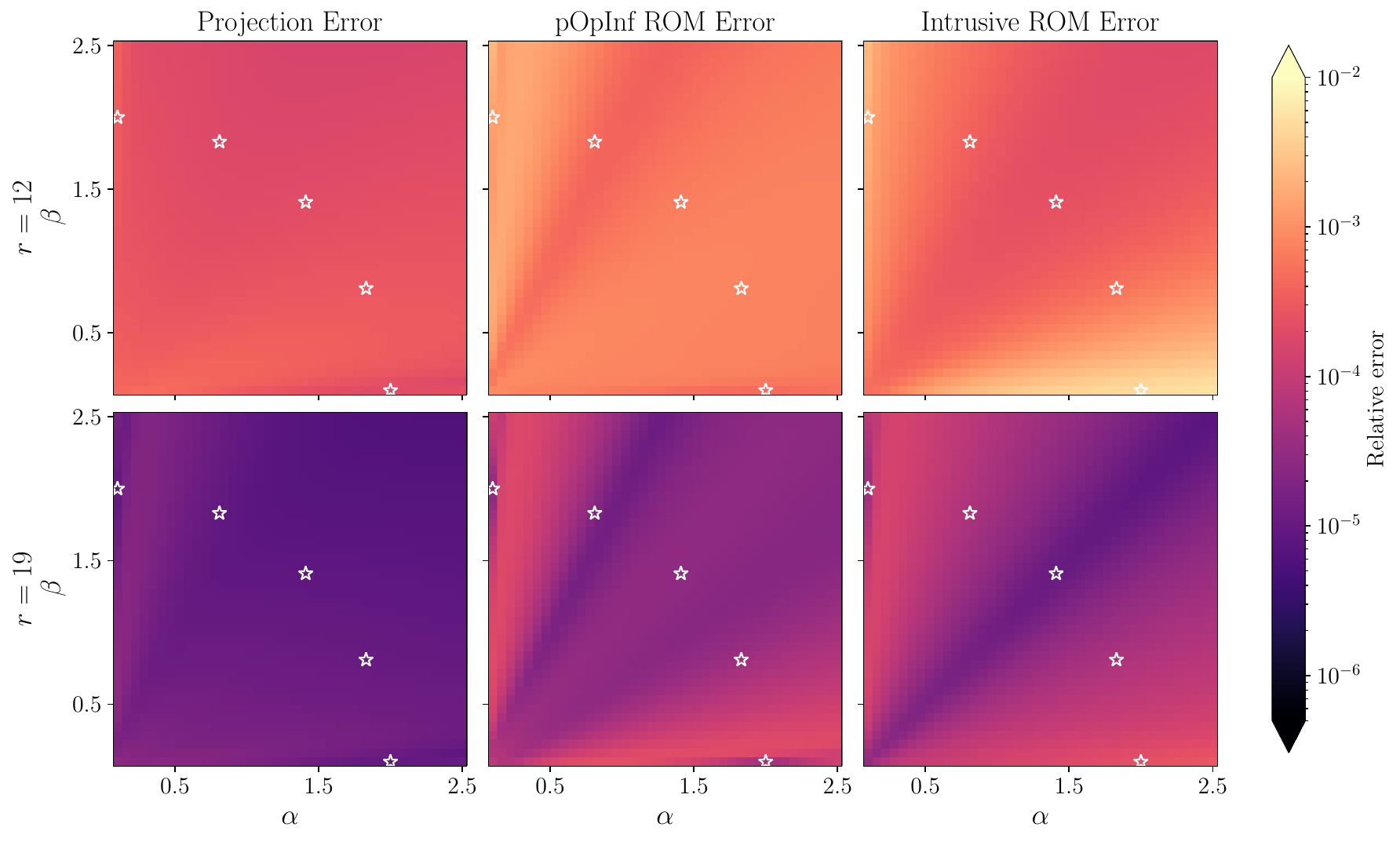}
    \vspace{-0.75cm}
    \caption{Relative $L^2$ projection errors (left), pOpInf ROM errors (center), and intrusive ROM errors (right)
    over the parameter domain $\mathcal{P}$ for the heat problem \cref{eq:heat-pde}--\cref{eq:heat-bcs}. Both ROMs use the same basis with $r = 12$ (top) or $r = 19$ (bottom) basis functions. The parameter samples used to generate the training set are marked as stars.}
    \label{fig:heat:param-errors}
\end{figure}

We compute $\widehat{\mathbf{O}}$ using \cref{alg:OpInf-reg-multi} with the $s = 5$ training parameter values and without any stability parameter values ($\bar{s} = 0$). Since \cref{eq:heat-ode} is a linear system, the regularization is parameterized by a single hyperparameter $\lambda_{1}$, hence the minimization~\cref{eq:opinf-reg} for this problem is given by
\begin{align*}
\min_{\widehat{\mathbf{O}}}\left\|
        \mathbf{D}\widehat{\mathbf{O}}^{\top} - \mathbf{R}^{\top}
    \right\|_{F}^{2}
    + \lambda_{1}^{2}\left(
        \|\widehat{\mathbf{A}}^{(1)}\|_{F}^{2} + \|\widehat{\mathbf{A}}^{(2)}\|_{F}^{2}
    \right).
\end{align*}
Here, the time derivative data $\mathbf{R}$ is estimated by first-order backward differences of the training states. The resulting ROM is integrated using the same implicit Euler scheme as the full-order model. To evaluate the performance of the ROM in terms of the parameters, we discretize $\mathcal{P}$ in a $40\times 40$ uniform grid. For each $\mu$ in the grid, we compute full-order solutions $\mathbf{u}(t_{1};\mu),\ldots,\mathbf{u}(t_{K};\mu)\in\mathbb{R}^{N}$ and integrate the ROM to obtain reduced states $\widetilde{\mathbf{u}}(t_{1};\mu),\ldots,\widetilde{\mathbf{u}}(t_{K};\mu)\in\mathbb{R}^{r}$, then compute the relative $L^{2}$-norm error
\begin{align}
    \label{eq:error-reconstruction}
    &\frac{\left\|
        \mathbf{V}\widetilde{\mathbf{u}}(\cdot\,;\mu) - \mathbf{u}(\cdot\,;\mu)
    \right\|_{L^{2}([t_0,t_f])}}{\left\|
        \mathbf{u}(\cdot\,;\mu)
    \right\|_{L^{2}([t_0,t_f])}},
    &
    \left\|\mathbf{w}(\cdot)\right\|_{L^{2}([t_0,t_f])}
    = \left(\int_{t_0}^{t_f}\left\|\mathbf{w}(t)\right\|_{2}^{2}\:dt\right)^{1/2},
\end{align}
estimating the time integrals via the trapezoidal rule. \Cref{fig:heat:param-errors} shows the results for $r = 12$ and $r = 19$ and compares them to the relative projection error induced by the corresponding bases, given by
\begin{align}
    \label{eq:error-projection}
    \frac{\left\|
        \mathbf{u}(\cdot\,;\mu)
        - \mathbf{V}\mathbf{V}^{\top}\mathbf{u}(\cdot\,;\mu)
    \right\|_{L^{2}([t_0,t_f])}}{\left\|
        \mathbf{u}(\cdot\,;\mu)
    \right\|_{L^{2}([t_0,t_f])}}.
\end{align}
Throughout $\mathcal{P}$, the relative projection error is on the order of $10^{-4}$ for $r = 12$ and $10^{-5}$ for $r = 19$, which indicates that the training snapshots have sufficient information to represent the solution well for any $\mu\in\mathcal{P}$. The results in \cref{fig:heat:param-errors} highlight the ability of pOpInf ROMs to generalize beyond the training data; i.e., the ROMs perform well for $(\alpha,\beta)$ pairs away from the arc $\alpha^{2} + \beta^{2} = 4$ where the parameter samples lie.

\Cref{fig:heat:param-errors} also shows the error of the ROM based on intrusive Galerkin projection,
\begin{align*}
    \frac{\mathrm{d}}{\mathrm{d}t}\widehat{\mathbf{u}}(t;\mu)
    &= \left(
        \alpha\bar{\mathbf{A}}^{(1)} + \beta\bar{\mathbf{A}}^{(2)}
    \right)\widehat{\mathbf{u}}(t;\mu),
\end{align*}
where the entries of $\bar{\mathbf{A}}^{(1)}$ and $\bar{\mathbf{A}}^{(1)}$ are computed as $[\bar{\mathbf{A}}^{(p)}]_{ij} = \left\langle v_i, \mathcal{A}^{(p)}\left(v_{j}\right)\right\rangle$, $p = 1,2$, as in \cref{eq:inner_product}. Note that this requires explicit access to the operators $\mathcal{A}^{(1)}$ and $\mathcal{A}^{(2)}$. The performance of the pOpInf ROMs---which are constructed without access to $\mathcal{A}^{(1)}$ and $\mathcal{A}^{(2)}$---is highly comparable to the performance of the intrusive ROMs. In particular, each ROM performs well near the line $\alpha = \beta$ and worse near the parameter domain boundaries $\alpha = 0.1$ and $\beta = 0.1$, with errors less than $0.6\%$ for $r = 12$ and $0.03\%$ for $r = 19$ throughout $\mathcal{P}$ (see \cref{table:heat-errors}). Using 100 random samples in $\mathcal{P}$, the average CPU time to integrate the full-order model is $\sim 0.087$~s, while the ROMs of size $r = 12$ integrate in $\sim 0.013$~s. Hence the computational speedup factor is $0.087 / 0.013 \approx 6.7$ times. The full-order solver takes advantage of sparsity in a large system, while the ROM utilizes a small but dense system.

\begin{table}
\footnotesize
\caption{Statistics for the relative errors displayed in \cref{fig:heat:param-errors}.}
\label{table:heat-errors}
\begin{center}
\begin{tabular}{cc|ccc}
    & & Projection & pOpInf ROM & Intrusive ROM \\
    \hline
             & maximum & $4.716\times 10^{-4}$ & $2.161\times 10^{-3}$ & $5.520\times 10^{-3}$ \\
    $r = 12$ & median & $2.297\times 10^{-4}$ & $7.415\times 10^{-4}$ & $3.862\times 10^{-4}$ \\
             & minimum & $1.610\times 10^{-4}$ & $3.683\times 10^{-4}$ & $1.904\times 10^{-4}$ \\
    \hline
             & maximum & $2.659\times 10^{-5}$ & $2.232\times 10^{-4}$ & $2.767\times 10^{-4}$ \\
    $r = 19$ & median & $1.034\times 10^{-5}$ & $2.915\times 10^{-5}$ & $4.386\times 10^{-5}$ \\
             & minimum & $5.895\times 10^{-6}$ & $1.212\times 10^{-5}$ & $7.643\times 10^{-6}$ \\
\end{tabular}
\end{center}
\end{table}

\subsection{FitzHugh--Nagumo System}
\label{subsec:numerical_fhn}
\begin{figure}[t]
    \centering
    \includegraphics[width=\textwidth]{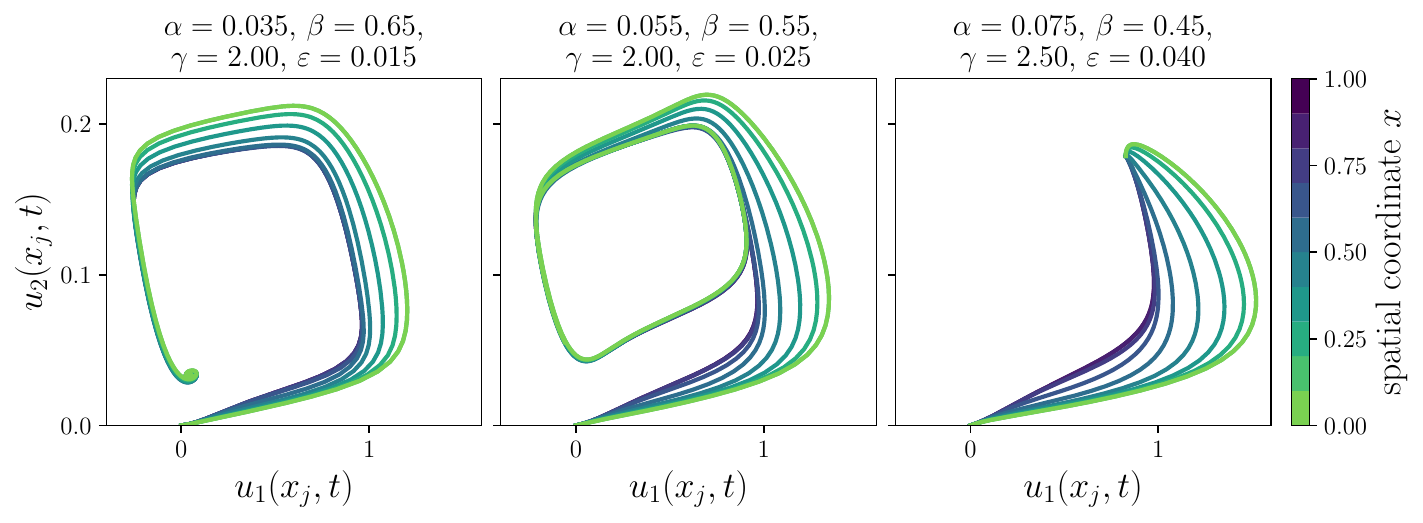}
    \vspace{-0.75cm}
    \caption{Phase portraits of training trajectories for the FitzHugh--Nagumo system \cref{eq:fhn-system}--\cref{eq:fhn-system2}, traced out at various points in the spatial domain. The center trajectory has a limit cycle, while the trajectories on the left and right converge to a single point. Small $\varepsilon$ values de-emphasize the diffusion term and drive the system toward spatial homogeneity; larger $\varepsilon$ values result in more variation across the spatial domain but which decreases with time.}
    \label{fig:fhn-training-data}
\end{figure}

The neuron model \cref{eq:fhn-system}--\cref{eq:fhn-system2} introduced in \cref{example:fh-n} features a four-dimensional parameter space. We generate training data at each of the $504 = 6\times 6 \times 2 \times 7$ unique parameter realizations $\mu = (\alpha,\beta,\gamma,\varepsilon)$ for
\begin{align}
    \tag{training set}
    \begin{aligned}
    \alpha &\in \{0.025, 0.035, \ldots, 0.075\},
    &
    \beta &\in \{0.25, 0.35, \ldots, 0.75\},
    \\
    \gamma &\in \{2.0, 2.5\},
    &
    \varepsilon &\in \{0.010, 0.015, \ldots, 0.040\}.
    \end{aligned}
\end{align}
For each parameter realization $\mu$ in the training set, we solve \cref{eq:fhn-system}--\cref{eq:fhn-system2} by discretizing the domain $\Omega = (0,1)$ with $N_{x} = 512$ spatial points and approximating the differential term with central finite differences. The total spatial dimension of this full-order model is thus $N = 2N_{x} = 1024$. Because the FitzHugh--Nagumo system is stiff, the spatially discretized equations benefit from a time integrator with an adaptive time step \cite{curtiss1952stiff,soderlind2002adaptive}. We make use of the implicit fifth-order Runge--Kutta Radau IIA method with adaptive time stepping \cite{wanner1996odes}, available in Python as \texttt{scipy.integrate.Radau} \cite{2020SciPy-NMeth}, with absolute and relative tolerances set to $10^{-6}$. The solution is computed at equally spaced times over the time domain $[t_{0}, t_{f}] = [0, 4]$ with spacing $\delta t = 10^{-3}$, after which the time derivatives of the states are estimated with sixth-order finite differences. The results are recorded for every tenth snapshot-derivative pair, resulting in $K = 400$ snapshots per training parameter realization with spacing $\delta t = 10^{-2}$. \Cref{fig:fhn-training-data} shows the phase plot of $u_{1}$ against $u_{2}$ at multiple spatial coordinate values for three of the training parameter samples, demonstrating that the system exhibits a diverse range of dynamical behaviors as the parameters are varied. To evaluate ROM performance with respect to the parameters, we also solve the full-order model at the $10{,}749 = 11\times 11\times 3\times 31 - 504$ additional parameter realizations $\mu = (\alpha,\beta,\gamma,\varepsilon)$ with
\begin{align}
    \tag{testing set}
    \begin{aligned}
    \alpha &\in \{0.025, 0.030, \ldots, 0.075\},
    &
    \beta &\in \{0.25, 0.30, \ldots, 0.75\},
    \\
    \gamma &\in \{2.00, 2.25, 2.50\},
    &
    \varepsilon &\in \{0.010, 0.011, \ldots, 0.040\}.
    \end{aligned}
\end{align}
Parameter realizations in the training set are not included in the testing set.

\begin{figure}[t]
\begin{minipage}[b]{0.52\textwidth}
    \begin{figure}[H]
        \centering
        \includegraphics[width=\textwidth]{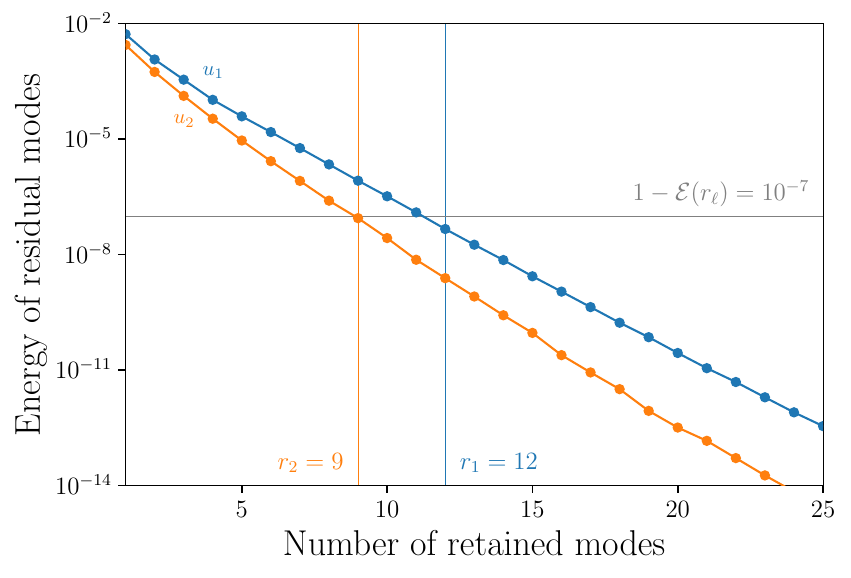}
    \end{figure}
\end{minipage}
\hfill
\begin{minipage}[b]{0.45\textwidth}
    \centering
    {\footnotesize
        \begin{center}
        \begin{tabular}{c|cc}
            $1 - \mathcal{E}(r_{\ell})$ & $r_{1}$ & $r_{2}$
            \\ \hline
            $10^{-3}$  & $ 3$ & $ 2$ \\
            $10^{-4}$  & $ 4$ & $ 4$ \\
            $10^{-5}$  & $ 7$ & $ 5$ \\
            $10^{-6}$  & $ 9$ & $ 7$ \\
            $10^{-7}$  & $12$ & $ 9$ \\
            $10^{-8}$  & $14$ & $11$ \\
            $10^{-9}$  & $17$ & $13$ \\
            $10^{-10}$ & $19$ & $15$ \\
            $10^{-11}$ & $22$ & $17$ \\
            $10^{-12}$ & $24$ & $19$ \\
        \end{tabular}
        \end{center}
    }
    \vfill
\end{minipage}
\vspace{-.075cm}
\caption{Decay of the residual energy in the training data for $u_1$ and $u_2$ in the FitzHugh--Nagumo problem \cref{eq:fhn-system}--\cref{eq:fhn-system2} (left) and the corresponding selected basis sizes (right). Demanding that $1 - \mathcal{E}(r_\ell) < 10^{-7}$ requires $r_{1} = 12$ and $r_{2} = 9$ POD modes for $u_{1}$ and $u_{2}$, respectively.}
\label{fig:fhn-residual-energy}
\end{figure}

\begin{figure}[t]
    \centering
    \includegraphics[width=\textwidth]{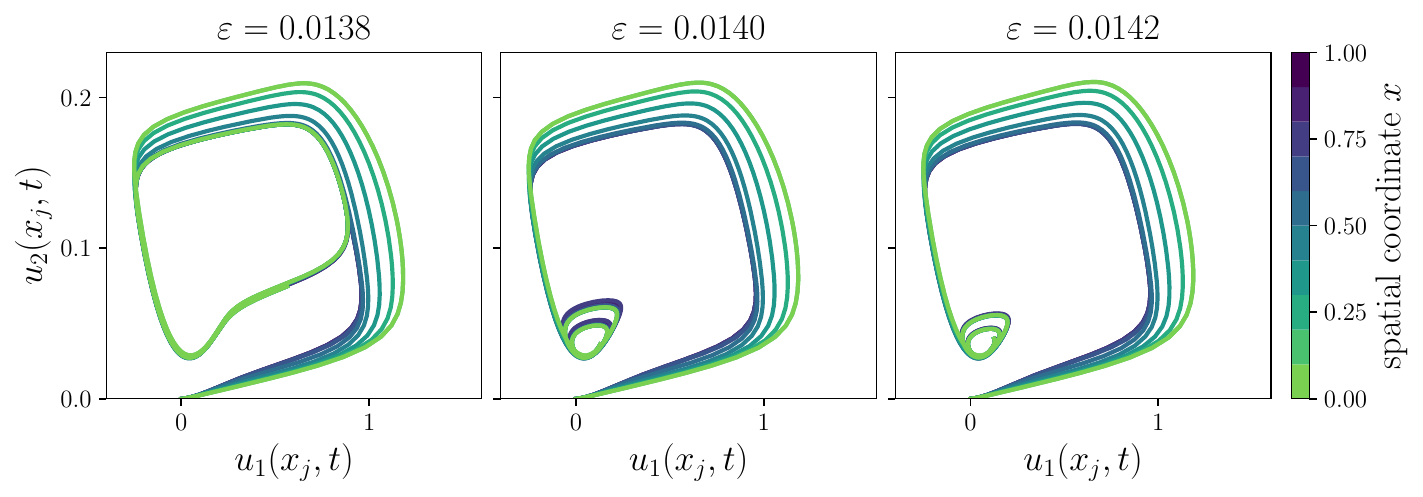}
    \vspace{-0.75cm}
    \caption{Phase portraits of trajectories for the FitzHugh--Nagumo system \cref{eq:fhn-system}--\cref{eq:fhn-system2}, traced out at various points in the spatial domain, at the parameter realizations $\alpha = 0.035$, $\beta = 0.75$, $\gamma = 2.5$, and three choices of $\varepsilon$. Near the Hopf bifurcation, small changes in $\varepsilon$ result in rapid changes in the system behavior.}
    \label{fig:fhn-phase-transition}
\end{figure}

The FitzHugh--Nagumo system \cref{eq:fhn-system}--\cref{eq:fhn-system2} exhibits a rapid singular Hopf bifurcation with respect to $\varepsilon$ wherein the limit cycle collapses to a stable fixed point as $\varepsilon$ increases \cite{baer1986singular}. Near this transition, the full-order model and ROM solutions are highly sensitive to $\varepsilon$ and the time integration scheme (see \cref{fig:fhn-phase-transition}). Including such parameter realizations in the training and testing sets makes it difficult to assess ROM accuracy. For each parameter realization $\mu = (\alpha, \beta, \gamma, \varepsilon)$ in the training and testing sets, we compare the corresponding full-order model solution to the solutions at $\mu_{\varepsilon-} = (\alpha, \beta, \gamma, \varepsilon - 0.001)$ and $\mu_{\varepsilon+} = (\alpha, \beta, \gamma, \varepsilon + 0.001)$. If the relative difference between neighboring solutions exceeds $50\%$, that is, if
\begin{align*}
    \max\left\{
    \frac{\left\|
        \mathbf{u}(\cdot\,;\mu)
        - \mathbf{u}(\cdot\,;\mu_{\varepsilon-})
    \right\|_{L^{2}([t_0,t_f])}}{\left\|
        \mathbf{u}(\cdot\,;\mu)
    \right\|_{L^{2}([t_0,t_f])}},
    \
    \frac{\left\|
        \mathbf{u}(\cdot\,;\mu)
        - \mathbf{u}(\cdot\,;\mu_{\varepsilon+})
    \right\|_{L^{2}([t_0,t_f])}}{\left\|
        \mathbf{u}(\cdot\,;\mu)
    \right\|_{L^{2}([t_0,t_f])}},
    \right\}
    > 0.5,
\end{align*}
then we remove $\mu$ from the data. This prompts us to remove two parameter realizations from the training set ($\sim 0.4\%$ of the original training set) and eighty-nine parameter realizations from the testing set ($\sim 0.8\%$ of the original testing set). Hence the size of the final training set is $s = 504 - 2 = 502$ and the size of the final testing set is $10{,}749 - 89 = 10{,}660$, with size ratio $10{,}660 / 502 \approx 21.24$.

We select the number of POD modes $r_{1}$ and $r_{2}$ for the variables $u_{1}$ and $u_{2}$, respectively, based on the residual energy $1 - \mathcal{E}(r_{\ell})$ of the training data, $\ell=1,2$. For each of the resulting $(r_{1},r_{2})$ pairs shown in \Cref{fig:fhn-residual-energy}, we learn a pOpInf ROM via \cref{alg:OpInf-reg-multi} with $\bar{s} = 0$. To simplify the hyperparameter search, we regularize only the quadratic and cubic terms, so that the regularized pOpInf data residuals are
\begin{align*}
    \min_{\widehat{\mathbf{O}}_{1}}
    &\left\|
        \mathbf{D}_{1}\widehat{\mathbf{O}}_{1}^{\top} - \mathbf{R}_{1}^{\top}
    \right\|_{F}^{2}
    + \lambda_{2}\left\|\widehat{\mathbf{H}}_{1,11}^{(1)}\right\|_{F}^{2}
    + \lambda_{3}\left\|\widehat{\mathbf{G}}_{1,111}^{(1)}\right\|_{F}^{2},
    &
    \min_{\widehat{\mathbf{O}}_{2}}
    &\left\|
        \mathbf{D}_{2}\widehat{\mathbf{O}}_{2}^{\top} - \mathbf{R}_{2}^{\top}
    \right\|_{F}^{2}.
\end{align*}
Regularization of the additional terms ($\lambda_{1}$ in \cref{eq:regularization-explicit}) was observed to have a marginal effect on the results. For each chosen basis size, we also compute a ROM based on intrusive projection for comparison with the pOpInf ROM. The intrusive ROM is derived explicitly without any need for approximation of the nonlinear terms via DEIM (see \cref{remark:FHN-DEIM}). The ROMs are integrated for each parameter quadruple $\mu = (\alpha, \beta, \gamma, \varepsilon)$ in the testing set using the same adaptive time integration scheme as the full-order model. Both ROMs are stable throughout the testing set with the exception of the pOpInf ROM with basis sizes $r_{1} = 7$ and $r_{2} = 5$, for which the time integrator diverges at three testing parameter realizations. For this case, we repeat \cref{alg:OpInf-reg-multi} with the same training data as before but now using these $\bar{s} = 3$ testing parameter realizations as the stability parameter inputs $\bar{\mu}_{1},\bar{\mu}_{2},\bar{\mu}_{3}$. The resulting ROM is then stable throughout the entire testing set.

\begin{figure}
    \centering
    \includegraphics[width=\textwidth]{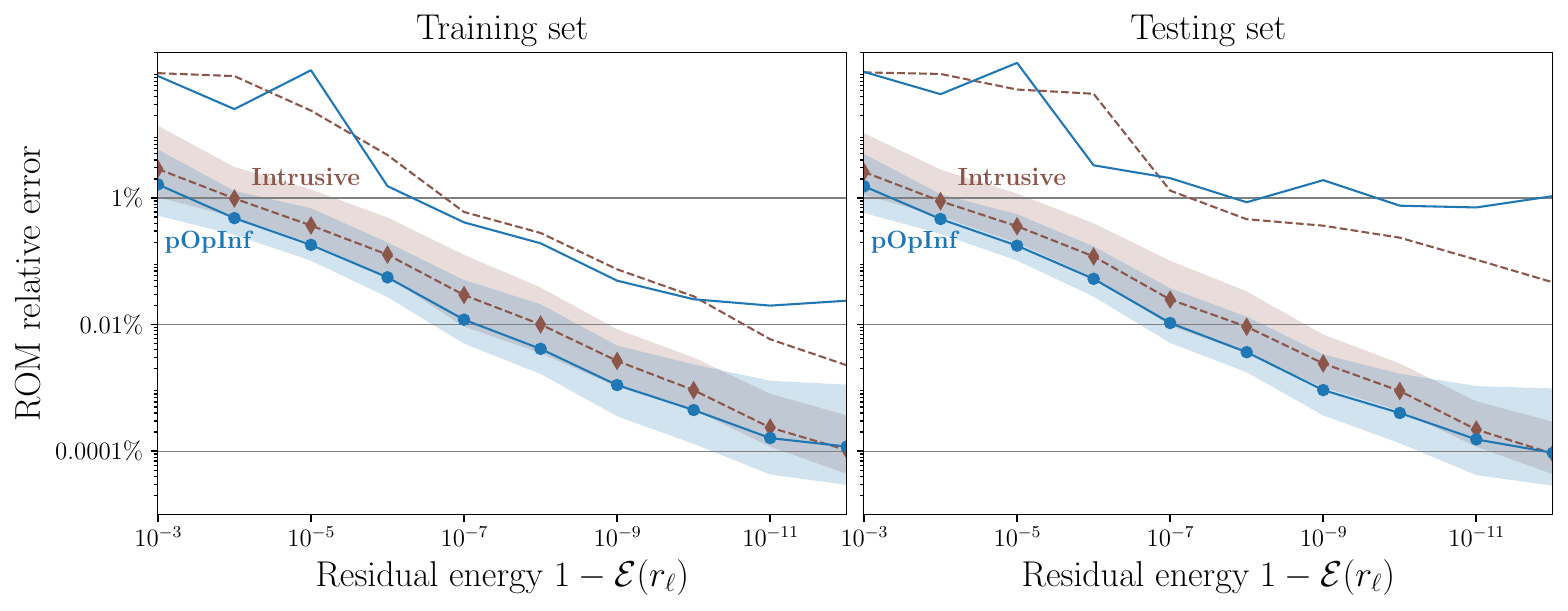}
    \vspace{-0.75cm}
    \caption{Relative errors (calculated over space and time) of the learned pOpInf ROM and the intrusive ROM for the FitzHugh--Nagumo system. As residual energy decreases from left to right, the sizes of the underlying POD bases increase according to \Cref{fig:fhn-residual-energy}. The shaded regions show the $10\%$--$90\%$ interdecile range of the error across all training samples (left) or testing samples (right), with the corresponding median and maximum errors denoted by the lines for the pOpInf ROMs and dashed lines for the intrusive ROMs.}
    \label{fig:fhn-basisVerror}
\end{figure}

\begin{figure}
    \centering
    \includegraphics[width=\textwidth]{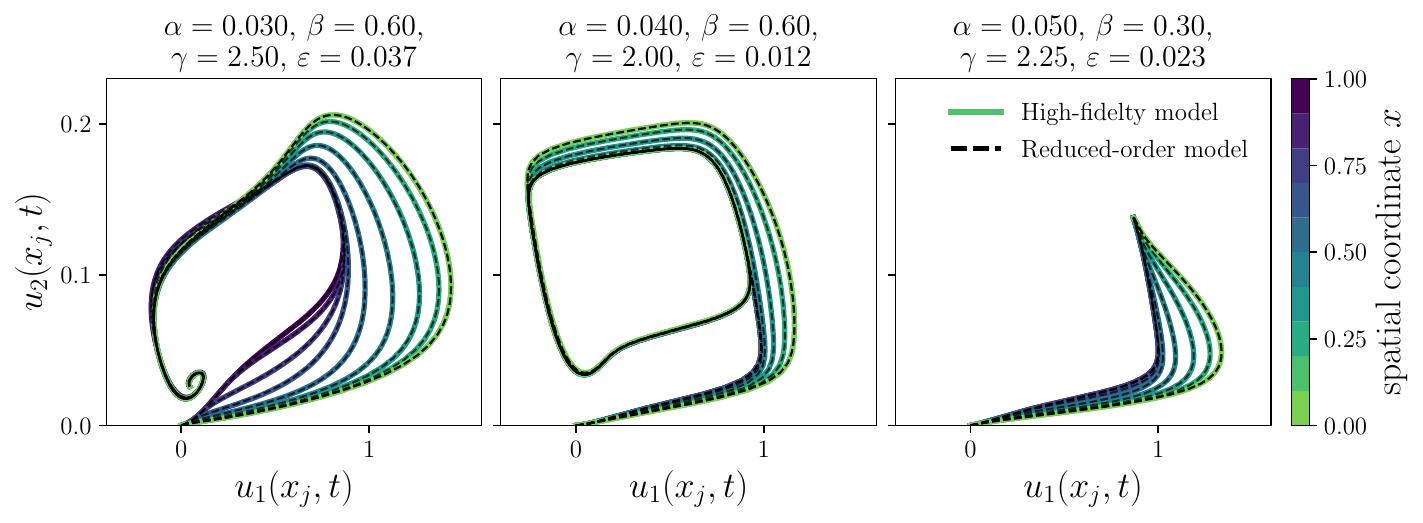}
    \vspace{-0.75cm}
    \caption{Phase portraits of testing trajectories for the FitzHugh--Nagumo problem \cref{eq:fhn-system}--\cref{eq:fhn-system2}, traced out at various points in the spatial domain. The solid lines are the full-order trajectories, and the dashed lines are the outputs of the learned pOpInf ROM with $r_{1} = 12$ and $r_{2} = 9$ POD modes. The total ROM relative errors in space and time are, from left to right, $0.028\%$, $0.190\%$, and $0.007\%$; the median relative error on the testing set for $r_1=12$ and $r_2=9$ is $0.011\%$.}
    \label{fig:fhn-fomromphase}
\end{figure}

The relative error compared to the full-order model is computed as in \cref{eq:error-reconstruction}, yielding a single error value for each parameter realization in the training and testing sets. \Cref{fig:fhn-basisVerror} shows the $10\%$ quantile, median, $90\%$ quantile, and maximum of the relative errors for each ROM. The median relative error is similar in the training and testing sets and decreases steadily as the basis sizes increase. The pOpInf ROMs generally outperform the intrusive ROMs in terms of median error. This is unsurprising for the training set since the pOpInf ROMs minimize a residual driven by the training data, whereas the intrusive ROMs only see the data through the construction of the POD bases. The tight interdecile range shows that the ROM error is mostly consistent throughout the parameter space; however, as the basis sizes increase the maximum pOpInf ROM error in both the training and testing sets eventually levels out while the maximum intrusive ROM error continues to decrease. For additional comparison, \Cref{fig:fhn-fomromphase} shows select trajectories of a single pOpInf ROM versus the full-order model. The reduced-order and full-order results are indistinguishable to the eye, which is representative of ROM performance throughout the testing set.

\begin{figure}
    \centering
    \includegraphics[width=\textwidth]{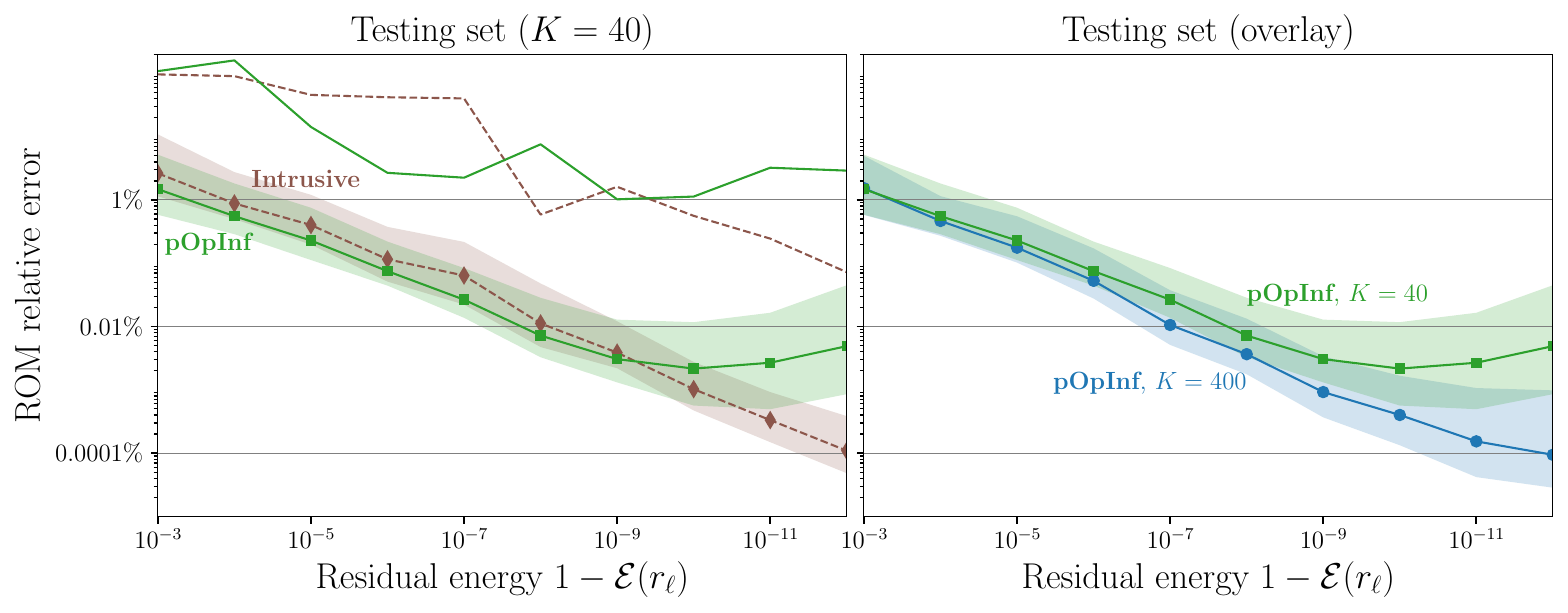}
    \vspace{-0.75cm}
    \caption{Relative errors on the testing set (calculated over space and time) of the learned pOpInf ROM for the FitzHugh--Nagumo system with only $K = 40$ snapshots per training parameter sample. The shaded regions show the $10\%$--$90\%$ interdecile range of the error; median and maximum errors are denoted by solid lines. On the left, a comparison to the corresponding intrusive ROM (dashed lines); on the right, a comparison to the pOpInf ROM with $K = 400$ snapshots per training parameter from \cref{fig:fhn-basisVerror}.}
    \label{fig:fhn-basisVerrorComparison}
\end{figure}

To probe the sensitivity of the method to the amount of training data, we repeat the numerical experiments by again estimating the time derivatives of the state with $\delta t = 10^{-3}$, then downsampling the snapshot data to only $K = 40$ snapshots per training parameter with uniform temporal spacing $\delta t = 10^{-1}$. This is equivalent to using only every tenth snapshot from the previous experiment. From this dataset we compute new POD bases and the corresponding intrusive ROMs. An initial application of \cref{alg:OpInf-reg-multi} with $\bar{s} = 0$ yields pOpInf ROMs that are stable except at three and ten testing parameter realizations for $(r_{1}, r_{2}) = (5, 4)$ and $(r_{1},r_{2}) = (21, 17)$, respectively. For these models, reapplying \cref{alg:OpInf-reg-multi} with these testing parameter realizations as the stability parameter inputs $\bar{\mu}_{1},\ldots,\bar{\mu}_{\bar{s}}$ ($\bar{s} = 3$ and $\bar{s} = 10$, respectively) yields pOpInf ROMs that are stable over the entire testing set. \Cref{fig:fhn-basisVerrorComparison} shows the relative errors of the new intrusive and pOpInf ROMs over the testing set and compares the errors of this $K = 40$ experiment with the errors from the original $K = 400$ experiment. The $K=40$ intrusive ROM error is similar to the $K = 400$ results shown in \Cref{fig:fhn-basisVerror}, but the pOpInf ROM error levels out near $\mathcal{E}(r_{\ell}) = 10^{-10}$, after which the intrusive ROM outperforms the pOpInf ROM. Though not shown, the same trend occurs for the error over the training set. With $K = 40$, the median pOpInf ROM error is consistently slightly higher than the $K = 400$ pOpInf ROM error, and the gap between the two errors increase with the basis sizes. For this problem, \cref{alg:OpInf-reg-multi} produces more robust and accurate ROMs when more training data are available.

\section{Conclusions}
\label{sec:conclusions}
This paper has proposed a non-intrusive parametric model reduction method for parameterized PDEs based on the Operator Inference framework. The approach eliminates the need for interpolation by explicitly embedding the polynomial affine-parametric structure of the PDE system into the resulting ROM and uses an optimization-based regularization strategy to ensure well-posedness in the learning problem. The parametric ROMs can later be used in outer-loop applications to expedite model evaluations for any choice of the parameters. The efficacy of the method has been demonstrated for two numerical examples: a heat equation with a two-dimensional parametric space, and the FitzHugh--Nagumo system with a four-dimensional parametric space. The resulting ROMs are capable of capturing the behavior of the PDE for parameters outside of the training set and perform favorably on average when compared to the intrusive ROM with the same affine structure.

It was shown in the FitzHugh--Nagumo example that the learned ROMs successfully capture the inherently different behaviors of the system that come with changes in the parameters (see \Cref{fig:fhn-fomromphase}). Yet, as with other data-driven approaches, the quality of inferred ROMs depends strongly on the training set, and one cannot expect a data-driven ROM to produce a particular dynamical behavior that differs wildly from the training data. An important future direction is therefore the automation of an efficient parameter sampling strategy for the offline stage. We further showed that the performance of pOpInf degrades slightly as data becomes more sparse in time, although the results can still be quite good as long as the time derivatives are estimated accurately. Further research is needed to address problems for which data are sparse in time and cannot be upsampled to produce accurate time derivative estimates via finite differences. Finally, as the framework proposed here relies on an affine structure in the parametric dependence, future work should address adapting this framework to non-affine parametric problems or to problems for which the parametric structure is unknown.

\section*{Acknowledgements}
This work has been supported in part by the US Department of Energy AEOLUS MMICC center under award DE-SC0019303, program manager W.~Spotz; by the Air Force Center of Excellence on Multi-Fidelity Modeling of Rocket Combustor Dynamics under award FA9550-17-1-0195; and by the US Department of Energy National Nuclear Security Administration under award DE-NA0003969. The authors also wish to thank Elizabeth Qian and Vincent Martinez for their insightful comments.

\appendix

\newpage
\section{Matricization of Tensors}
\label{appendix:kronecker}
Following the notation of \cite{kolda2009tensor}, for matrices $\mathbf{W}\in \mathbb{R}^{r\times s}$ and $\mathbf{Z}\in \mathbb{R}^{m\times n}$, let $\mathbf{W}\otimes\mathbf{Z}$ denote the Kronecker product \cite{kolda2009tensor,vanLoan2000kronecker}:
\begin{align*}
    \mathbf{W}\otimes \mathbf{Z}=\left[\begin{array}{ccc}
        w_{11} & \cdots & w_{1s} \\
        \vdots & \ddots & \vdots \\
        w_{r1} & \cdots & w_{rs} \\
    \end{array}\right] \otimes \mathbf{Z}
    := \left[\begin{array}{ccc}
        w_{11}\mathbf{Z} & \cdots & w_{1s}\mathbf{Z} \\
        \vdots & \ddots & \vdots \\
        w_{r1}\mathbf{Z} & \cdots & w_{rs}\mathbf{Z} \\
    \end{array}\right]\in \mathbb{R}^{rm\times sn},
\end{align*}
where $w_{ij}$ is the component from the $i$th row and $j$th column of $\mathbf{W}$. The definition applies to vectors by setting $s = n = 1$. For $\mathbf{Z}\in\mathbb{R}^{m\times s}$ (i.e., $n = s$), define $\mathbf{W}\odot\mathbf{Z}$ to be the Khatri-Rao product, i.e., the column-wise Kronecker product \cite{KhatriRoa1968product,kolda2009tensor}:
\begin{align*}
    \mathbf{W}\odot \mathbf{Z}
    :=
    \left[\begin{array}{c|c|c|c}
        \mathbf{w}_{1} \otimes \mathbf{z}_{1}
        &\mathbf{w}_{2} \otimes \mathbf{z}_{2}
        & \cdots &
        \mathbf{w}_{s} \otimes \mathbf{z}_{s}
    \end{array}\right]\in \mathbb{R}^{rm\times s},
\end{align*}
where $\mathbf{w}_i$ and $\mathbf{z}_i$ are the $i$th columns of $\mathbf{W}$ and $\mathbf{Z}$, respectively.

Let $\mathbf{w} = [w_1~\cdots~w_{r}]^\top \in \mathbb{R}^{r}$. As the product $\mathbf{w}\odot\mathbf{w} = \mathbf{w}\otimes\mathbf{w}$ has redundant terms (for instance, $w_1w_2 = w_2w_1$ appears twice), we introduce compact second- and third-order Khatri-Rao products, defined as
\begin{align*}
    \mathbf{w}\, \widehat{\odot}\, \mathbf{w}
    &:= \left[\begin{array}{c}
        w_{1}^2
        \\
        w_{2}(\mathbf{w}_{1:2})
        \\ \vdots \\
        w_{r}(\mathbf{w}_{1:r})
    \end{array}\right] \in \mathbb{R}^{\binom{r+1}{2}},
    &
    \mathbf{w}\, \widehat{\odot}\, \mathbf{w}\, \widehat{\odot}\, \mathbf{w}
    &:= \left[\begin{array}{c}
        w_{1}^3
        \\
        w_{2}(\mathbf{w}_{1:2}\, \widehat{\odot}\, \mathbf{w}_{1:2})
        \\ \vdots \\
        w_{r}(\mathbf{w}_{1:r}\, \widehat{\odot}\, \mathbf{w}_{1:r})
    \end{array}\right] \in \mathbb{R}^{\binom{r+2}{3}},
\end{align*}
where $\mathbf{w}_{1:i} = [w_1~\cdots~w_{i}]^\top\in\mathbb{R}^{i}$ contains the first $i$ entries of $\mathbf{w}$. Each entry of $\mathbf{w}\,\widehat{\odot}\,\mathbf{w}$ is a product of 2 entries of $\mathbf{w}$, and each product appears exactly once; in other words, $\mathbf{w}\,\widehat{\odot}\,\mathbf{w}$ contains the unique terms of $\mathbf{w}\mathbf{w}^\top$, the 2-fold tensorization of $\mathbf{w}$ \cite{kolda2009tensor,OPBWN2018matrixcompletion}. Similarly, each entry of $\mathbf{w}\,\widehat{\odot}\,\mathbf{w}\,\widehat{\odot}\,\mathbf{w}$ is a unique product of 3 entries of $\mathbf{w}$. For matrices, the definition applies column-wise:
\begin{align*}
    \left[\begin{array}{c|c|c}
        \mathbf{w}_{1} & \cdots & \mathbf{w}_{s}
    \end{array}\right]\,\widehat{\odot}\, \left[\begin{array}{c|c|c}
        \mathbf{w}_{1} & \cdots & \mathbf{w}_{s}
    \end{array}\right]
    =
    \left[\begin{array}{c|c|c}
        \mathbf{w}_{1}\,\widehat{\odot}\,\mathbf{w}_{1}
        & \cdots &
        \mathbf{w}_{s}\,\widehat{\odot}\,\mathbf{w}_{s}
    \end{array}\right].
\end{align*}

With this notation, any linear combination of products of two entries of $\mathbf{w}$ can be represented by the product
\begin{align*}
    \mathbf{z}^\top\left(\mathbf{w} \,\widehat{\odot}\, \mathbf{w}\right)
    &= z_1 w_1^2 + z_2 w_1 w_2 + z_3 w_2^2 + z_4 w_1 w_3 + \cdots,
    &
    \mathbf{z}
    &= \left[\begin{array}{c} z_1 \\ z_2 \\ \vdots \end{array}\right]
    \in\mathbb{R}^{\binom{r+1}{2}}.
\end{align*}
This observation allows us to convert \cref{eq:inner_product} to a compact matrix product representation. Specifically, we define $\widehat{\mathbf{H}}\in\mathbb{R}^{r \times \binom{r+1}{2}}$ to be the matrix such that the $i$th component of the product $\widehat{\mathbf{H}}\left(\widehat{\mathbf{u}}\,\widehat{\odot}\,\widehat{\mathbf{u}}\right)$ is given by
\begin{align*}
    [\widehat{\mathbf{H}}\left(\widehat{\mathbf{u}}\,\widehat{\odot}\,\widehat{\mathbf{u}}\right)]_{i}
    := \sum_{j=1}^{r} \sum_{l=1}^{r} \left\langle v_i,
        \mathcal{H}\left(v_{j},v_{l};\mu\right)
    \right\rangle \hat{u}_{j} \hat{u}_{l}.
\end{align*}

\newpage
\section{Technical Lemma}
\label{appendix:aux_lemma}
The following lemma supports \Cref{thm:rank-deficiencies}. Here, $\mathbf{0}_{k}\in\mathbb{R}^{k}$ is the vector of $k$ zeros and $\mathbf{1}_{k}\in\mathbb{R}^{k}$ denotes the vector of $k$ ones.
\begin{lemma}
\label{lemma:kronecker-structure}
Let $\mathbf{y}_{1},\ldots,\mathbf{y}_{s} \in \mathbb{R}^{q}$ and $\mathbf{Z}_{1},\ldots,\mathbf{Z}_{s} \in \mathbb{R}^{k\times r}$. Consider the matrix
\begin{align*}
    \mathbf{W} := \left[\begin{array}{c}
        \mathbf{y}_{1}^\top \otimes \mathbf{Z}_{1}
        \\ \vdots \\
        \mathbf{y}_{s}^\top \otimes \mathbf{Z}_{s}
    \end{array}\right] \in \mathbb{R}^{sk \times qr}.
\end{align*}
If either of the matrices
\begin{align*}
    \mathbf{Y}
    &:= \left[\begin{array}{c}
        \mathbf{y}_{1}^\top \\ \vdots \\ \mathbf{y}_{s}^\top
    \end{array}\right] \in \mathbb{R}^{s \times q},
    &
    \mathbf{Z}
    &:= \left[\begin{array}{c}
        \mathbf{Z}_{1} \\ \vdots \\ \mathbf{Z}_{s}
    \end{array}\right] \in \mathbb{R}^{sk \times r}
\end{align*}
do not have full column rank, then neither does $\mathbf{W}$. Conversely, if $\mathbf{Y}$ and \underline{each} $\mathbf{Z}_{1},\ldots,\mathbf{Z}_{s}$ have full column rank, then so does $\mathbf{W}$.
\begin{proof}
Assume that $\mathbf{Y}$ does not have full column rank. Then there exists a nonzero vector $\boldsymbol{\alpha}\in\mathbb{R}^{q}$ such that $\mathbf{Y}\boldsymbol{\alpha} = \mathbf{0}_{s}$, that is, $\mathbf{y}_{i}^\top \boldsymbol{\alpha} = 0$ for $i = 1,\ldots,s$. Using the mixed-product property of the Kronecker product,
\begin{align*}
    \left(\mathbf{y}_{i}^\top \otimes \mathbf{Z}_{i}\right)
    \left(\boldsymbol{\alpha} \otimes \mathbf{1}_{r}\right)
    = \left(\mathbf{y}_{i}^\top \boldsymbol{\alpha}\right)
    \otimes
    \left(\mathbf{Z}_{i} \mathbf{1}_{r}\right)
    = 0
    \otimes
    \left(\mathbf{Z}_{i} \mathbf{1}_{r}\right)
    = \mathbf{0}_{k}
\end{align*}
for $i=1,\ldots,s$. Then $\mathbf{W}\left(\boldsymbol{\alpha} \otimes \mathbf{1}_{r}\right) = \mathbf{0}_{sk}$. But $\boldsymbol{\alpha} \otimes \mathbf{1}_{r}$ is a nonzero vector, which implies that the columns of $\mathbf{W}$ are linearly dependent.

Next, suppose that $\mathbf{Z}$ does not have full column rank. Then $\mathbf{Z}\boldsymbol{\beta} = \mathbf{0}_{sk}$ for some nonzero vector $\boldsymbol{\beta} \in \mathbb{R}^{r}$, implying $\mathbf{Z}_{i}\boldsymbol{\beta} = \mathbf{0}_{k}$ for $i=1,\ldots,s$. Then
\begin{align*}
    \left(\mathbf{y}_{i}^\top \otimes \mathbf{Z}_{i}\right)
    \left(\mathbf{1}_{q} \otimes \boldsymbol{\beta}\right)
    = \left(\mathbf{y}_{i}^\top \mathbf{1}_{q}\right)
    \otimes
    \left(\mathbf{Z}_{i} \boldsymbol{\beta}\right)
    = \left(\mathbf{y}_{i}^\top \mathbf{1}_{q}\right)
    \otimes
    \mathbf{0}_{k}
    = \mathbf{0}_{k}
\end{align*}
for $i=1,\ldots,s$, so that $\mathbf{W}\left(\mathbf{1}_{q} \otimes \boldsymbol{\beta}\right) = \mathbf{0}_{sk}$. Since $\mathbf{1}_{q} \otimes \boldsymbol{\beta}$ is a nonzero vector, $\mathbf{W}$ does not have full column rank.

For the converse statement, assume that $\mathbf{Y}$ and $\mathbf{Z}_{1},\ldots,\mathbf{Z}_{s}$ each have full column rank, and suppose
$
    \boldsymbol{\gamma}
    = [
    \boldsymbol{\gamma}_{1}^\top~\cdots~\boldsymbol{\gamma}_{q}^\top
    ]^\top\in\mathbb{R}^{qr}
$, $\boldsymbol{\gamma}_{j} \in \mathbb{R}^{r}$, satisfies $\mathbf{W}\boldsymbol{\gamma} = \mathbf{0}_{sk}$, i.e., $(\mathbf{y}_{i}^\top\otimes\mathbf{Z}_{i})\boldsymbol{\gamma} = \mathbf{0}_{k}$ for $i=1,\ldots,s$. Denoting
$
    \boldsymbol{\Gamma}
    = [
        \boldsymbol{\gamma}_{1}~\cdots~\boldsymbol{\gamma}_{q}
    ]\in\mathbb{R}^{r\times q},
$
we have
\begin{align*}
    \mathbf{0}_{k}
    =
    (\mathbf{y}_{i}^\top \otimes \mathbf{Z}_{i})\boldsymbol{\gamma}
    = \sum_{j=1}^{q}y_{ij}\mathbf{Z}_{i}\boldsymbol{\gamma}_{j}
    = \mathbf{Z}_{i}\left(\sum_{j=1}^{q}y_{ij}\boldsymbol{\gamma}_{j}\right)
    = \mathbf{Z}_{i}
\boldsymbol{\Gamma}
    \mathbf{y}_{i},
\end{align*}
where $y_{ij}$ is the $j$th entry of $\mathbf{y}_{i}$. Since each $\mathbf{Z}_{i}$ has full column rank, it must be the case that $\boldsymbol{\Gamma}\mathbf{y}_{i} = \mathbf{0}_{r}$ for each $i=1,\ldots,s$, which in turn implies
$
    \mathbf{Y}
    \boldsymbol{\Gamma}^\top
    = \left(
    \boldsymbol{\Gamma}
    \mathbf{Y}^\top
    \right)^{\top}
    = \mathbf{0}_{s \times r}.
$
But $\mathbf{Y}$ having full column rank implies $\boldsymbol{\Gamma} = \mathbf{0}_{r \times q}$, hence $\boldsymbol{\gamma} = \mathbf{0}_{qr}$. Thus, the columns of $\mathbf{W}$ are linearly independent, so $\mathbf{W}$ has full column rank.
\end{proof}
\end{lemma}

\newpage
\section{General Construction for PDE Systems}
\label{appendix:systems}
We provide here a general construction for the affine Operator Inference problem for systems of PDEs to learn reduced-order models of the form \cref{eq:ode-system}--\cref{eq:state-vector-system}. The problem decouples into $d$ instances of \cref{eq:opinf-standard}, that is,
\begin{align*}
    &\min_{\widehat{\mathbf{O}}_{\ell}}\left\|
        \mathbf{D}_{\ell}\widehat{\mathbf{O}}_{\ell}^\top - \mathbf{R}_{\ell}^\top
    \right\|_{F}^2,
    &
    \ell &= 1,\ldots, d,
\end{align*}
where
\begin{subequations}
\begin{align*}
    \mathbf{D}_{\ell}&= \left[\begin{array}{c|ccc|cccc}
        \mathbf{D}_{c_{\ell}} &
        \mathbf{D}_{A_{\ell,1}}&
        \cdots &
        \mathbf{D}_{A_{\ell,d}} &
        \mathbf{D}_{H_{\ell,11}} &
        \mathbf{D}_{H_{\ell,12}} &
        \cdots &
        \mathbf{D}_{H_{\ell,dd}}
    \end{array}\right],
    \\
    \widehat{\mathbf{O}}_{\ell}& = \left[\begin{array}{c|c|c}
        \widehat{\boldsymbol{\mathcal{C}}}_{\ell} &
        \widehat{\boldsymbol{\mathcal{A}}}_{\ell,1}\ \cdots\ \widehat{\boldsymbol{\mathcal{A}}}_{\ell,d} &
        \widehat{\boldsymbol{\mathcal{H}}}_{\ell,11}\ \widehat{\boldsymbol{\mathcal{H}}}_{\ell,12}\ \cdots\ \widehat{\boldsymbol{\mathcal{H}}}_{\ell,dd}
    \end{array}\right],
    \\
    \mathbf{R}_{\ell} &= \left[\begin{array}{ccc}
        \dot{\widehat{\mathbf{U}}}_{\ell}(\mu_1)
        & \cdots &
        \dot{\widehat{\mathbf{U}}}_{\ell}(\mu_s)
    \end{array}\right] \in \mathbb{R}^{r_{\ell} \times sK},
\end{align*}
\end{subequations}
with
\begin{align*}
    \mathbf{D}_{c_\ell} & =\left[\begin{array}{c}
        \boldsymbol{\theta}_{c_{\ell}}(\mu_{1})\otimes\mathbf{1}_{K}\\
        \vdots\\
        \boldsymbol{\theta}_{c_{\ell}}(\mu_{s})\otimes\mathbf{1}_{K}
    \end{array}\right]\in \mathbb{R}^{K\times q_{c_{\ell}}},
    \\
    \mathbf{D}_{A_{\ell,m}} & =\left[\begin{array}{c}
        \boldsymbol{\theta}_{A_{\ell,m}}(\mu_{1})\otimes \widehat{\mathbf{U}}_m(\mu_{1})^\top\\
        \vdots\\
        \boldsymbol{\theta}_{A_{\ell,m}}(\mu_{s})\otimes \widehat{\mathbf{U}}_m(\mu_{s})^\top
    \end{array}\right]\in \mathbb{R}^{K\times q_{A_{\ell,m}}r_m},
    \\
    \mathbf{D}_{H_{\ell,mm}}& =\left[\begin{array}{c}
        \boldsymbol{\theta}_{H_{\ell,mm}}(\mu_{1})\otimes \left(\widehat{\mathbf{U}}_m(\mu_{1})\,\widehat{\odot}\,\widehat{\mathbf{U}}_m(\mu_{1})\right)^\top\\
        \vdots\\
        \boldsymbol{\theta}_{H_{\ell,mm}}(\mu_{s})\otimes  \left(\widehat{\mathbf{U}}_m(\mu_{s})\,\widehat{\odot}\,\widehat{\mathbf{U}}_m(\mu_{s})\right)^\top
    \end{array}\right]\in \mathbb{R}^{K\times q_{H_{\ell,mm}}r_m(r_m+1)/2},
    \\
    \mathbf{D}_{H_{\ell,mn}}& =\left[\begin{array}{c}
        \boldsymbol{\theta}_{H_{\ell,mn}}(\mu_{1})\otimes \left(\widehat{\mathbf{U}}_m(\mu_{1})\odot\widehat{\mathbf{U}}_n(\mu_{1})\right)^\top\\
        \vdots\\
        \boldsymbol{\theta}_{H_{\ell,mn}}(\mu_{s})\otimes  \left(\widehat{\mathbf{U}}_m(\mu_{s})\odot\widehat{\mathbf{U}}_n(\mu_{s})\right)^\top
    \end{array}\right]\in \mathbb{R}^{K\times q_{H_{\ell,mn}}r_mr_n}, \quad (n \neq m)
    \\
    \widehat{\boldsymbol{\mathcal{C}}}_{\ell} &=\left[\begin{array}{c}
        \widehat{\mathbf{c}}_{\ell}^{(1)}\ \cdots\ \widehat{\mathbf{c}}_{\ell}^{(q_{c_{\ell}})}
    \end{array}\right] \in \mathbb{R}^{r_{\ell} \times q_{c_{\ell}}}
    \\
    \widehat{\boldsymbol{\mathcal{A}}}_{\ell,m} & = \left[\begin{array}{c}
        \widehat{\mathbf{A}}_{\ell,m}^{(1)}\ \cdots\ \widehat{\mathbf{A}}_{\ell,m}^{(q_{A_{\ell,m}})}
    \end{array}\right] \in \mathbb{R}^{r_{\ell} \times q_{A_{\ell,m}}r_m}
    \\
    \widehat{\boldsymbol{\mathcal{H}}}_{\ell,mm} & = \left[\begin{array}{c}
        \widehat{\mathbf{H}}_{\ell,mm}^{(1)}\ \cdots\ \widehat{\mathbf{H}}_{\ell,mm}^{(q_{H_{\ell,mm}})}
    \end{array}\right] \in \mathbb{R}^{r_{\ell} \times q_{H_{\ell,mm}}r_m(r_m+1)/2}
    \\
    \widehat{\boldsymbol{\mathcal{H}}}_{\ell,mn} & = \left[\begin{array}{c}
        \widehat{\mathbf{H}}_{\ell,mn}^{(1)}\ \cdots\ \widehat{\mathbf{H}}_{\ell,mn}^{(q_{H_{\ell,mn}})}
    \end{array}\right] \in \mathbb{R}^{r_{\ell} \times q_{H_{\ell,mn}}r_mr_n}, \quad (m\neq n)
    \\
    \widehat{\mathbf{U}}_{\ell}(\mu_i) &= \left[\begin{array}{ccc}
        \widehat{\mathbf{u}}_{\ell}(t_{1};\mu_i) & \cdots & \widehat{\mathbf{u}}_{\ell}(t_{K};\mu_{i})
    \end{array}\right] \in \mathbb{R}^{r_{\ell} \times K},
    \\
    \dot{\widehat{\mathbf{U}}}_{\ell}(\mu_i) &= \left[\begin{array}{ccc}
        \frac{\textup{d}}{\textup{d}t}\widehat{\mathbf{u}}_{\ell}(t;\mu_i)\Bigr|_{t=t_{1}} & \cdots & \frac{\textup{d}}{\textup{d}t}\widehat{\mathbf{u}}_{\ell}(t;\mu_{i})\Bigr|_{t=t_{K}}
    \end{array}\right] \in \mathbb{R}^{r_{\ell} \times K},
    \\
    \boldsymbol{\theta}_{c_{\ell}}(\mu_{i}) &= \left[\begin{array}{ccc}
        \theta_{c_{\ell}}^{(1)}(\mu_{i}) & \cdots & \theta_{c_{\ell}}^{(q_{c_{\ell}})}(\mu_{i})
    \end{array}\right]\in\mathbb{R}^{1 \times q_{c_{\ell}}},
    \\
    \boldsymbol{\theta}_{A_{\ell,m}}(\mu_{i}) &= \left[\begin{array}{ccc}
        \theta_{A_{\ell,m}}^{(1)}(\mu_{i}) & \cdots & \theta_{A_{\ell,m}}^{(q_{A_{\ell,m}})}(\mu_{i})
    \end{array}\right]\in\mathbb{R}^{1 \times q_{A_{\ell,m}}},
    \\
    \boldsymbol{\theta}_{H_{\ell,mn}}(\mu_{i}) &= \left[\begin{array}{ccc}
        \theta_{H_{\ell,mn}}^{(1)}(\mu_{i}) & \cdots & \theta_{H_{\ell,mn}}^{(q_{H_{\ell,mn}})}(\mu_{i})
    \end{array}\right]\in\mathbb{R}^{1 \times q_{H_{\ell,mn}}}.
\end{align*}
In practice, this construction is often sparse due to the limited number of terms in the governing PDE.

\bibliographystyle{siamplain}
\bibliography{references}

\end{document}